%% file: main.tex
\pdfoutput=1
\documentclass[conference]{IEEEtran}
\usepackage{blindtext, graphicx}
\ifCLASSINFOpdf
\else
\fi
\usepackage{epsfig}
\usepackage{graphicx,xspace,xcolor,cite,url}
\usepackage{times}
\usepackage{algorithmic,algorithm}
\usepackage{xspace}
\usepackage{amssymb}
\usepackage{multirow}
\usepackage{booktabs}
\usepackage{listings}
\usepackage{comment}
\usepackage{fancybox}
\usepackage{xspace}
\usepackage{paralist}
\usepackage{caption}
\usepackage{subcaption}
\usepackage{colortbl}
\usepackage{flushend}     
\usepackage{dblfloatfix}    
\definecolor{Gray}{gray}{0.85}

\usepackage{array}
\newcolumntype{L}[1]{>{\raggedright\let\newline\\\arraybackslash\hspace{0pt}}m{#1}}
\newcolumntype{C}[1]{>{\centering\let\newline\\\arraybackslash\hspace{0pt}}m{#1}}
\newcolumntype{R}[1]{>{\raggedleft\let\newline\\\arraybackslash\hspace{0pt}}m{#1}}

\usepackage{amssymb}
\usepackage{pifont}
\newcommand{\cmark}{\ding{51}}%
\newcommand{\xmark}{\ding{55}}%

\newcommand{\sys}{dIFTTT\xspace}
\newcommand{\longsys}{Decoupled-IFTTT\xspace}

\usepackage[normalem]{ulem}
\newcommand{\eg}[0]{\textit{e.g.,}\xspace}
\newcommand{\ie}[0]{\textit{i.e.,}\xspace}
\newcommand{\etal}[0]{\textit{et al.}\xspace}
\newcommand{\xref}[1]{\S\ref{#1}}

\newcommand{\dcloud}[0]{\sys-Cloud\xspace}
\newcommand{\dclient}[0]{\sys-Client\xspace}
\newcommand{\dclients}[0]{\sys-Clients\xspace}

\IEEEoverridecommandlockouts

\input{stats}

\begin{document}
%
%
\title{\longsys: Constraining Privilege in Trigger-Action Platforms for the Internet of Things}

\author{\IEEEauthorblockN{Earlence Fernandes, Amir Rahmati}
\IEEEauthorblockA{University of Michigan}
\and
\IEEEauthorblockN{Jaeyeon Jung}
\IEEEauthorblockA{Samsung}
\and
\IEEEauthorblockN{Atul Prakash}
\IEEEauthorblockA{University of Michigan}
\thanks{Earlence Fernandes is now with the University of Washington. Contact him at earlence@cs.washington.edu}
\thanks{Amir Rahmati is now with Samsung Research.}
}

\maketitle

\input{abstract2}
\input{intro2}

\input{bg}
\input{empirical}

\input{lessons}
\input{defenses}

\input{eval}

\input{discussion}
\input{related}

\input{conclusion}
\input{ack}

\bibliographystyle{IEEEtranS}
\bibliography{reference}
\input{appa}

\end{document}

%% file: stats.tex
\newcommand{\totalChannels}{297\xspace} 
\newcommand{\connectedChannels}{170\xspace} 
\newcommand{\scrShotChannels}{128\xspace} 
\newcommand{\dateofdata}{April 14th, 2016\xspace} 
\newcommand{\iotChannels}{80\xspace} 
\newcommand{\defaultConnected}{40\xspace}
\newcommand{\overprivdChannelsAll}{18\xspace} 
\newcommand{\notOverprivdChannelsAll}{6\xspace} 
\newcommand{\avgOverprivAPIs}{26\xspace} 
\newcommand{\mobileChannels}{33\xspace}
\newcommand{\paywallChannels}{67\xspace}
\newcommand{\binaryOAuth}{122\xspace} 
\newcommand{\binarySingleTrigAct}{106\xspace} 
\newcommand{\fineOAuth}{4\xspace} 
\newcommand{\timedOAuth}{2\xspace} 
\newcommand{\goodinfo}{76\xspace}	
\newcommand{\okinfo}{25\xspace} 
\newcommand{\noinfo}{27\xspace} 

\newcommand{\numApisStudied}{941\xspace} 
\newcommand{\numChannelsStudiedAll}{24\xspace} 
\newcommand{\numChannelsStudiedIoT}{16\xspace} 
\newcommand{\measurableChannelsAll}{69\xspace} %
\newcommand{\recipeCoverageTrigger}{80.4\%\xspace} 
\newcommand{\shareCoverageTrigger}{86.5\%\xspace} 

\newcommand{\totalRecipes}{219,284\xspace} 
\newcommand{\numChannelsCouldNotAnalyze}{169\xspace} 
\newcommand{\defunctChannels}{29\xspace} 
\newcommand{\coveredRecipesTrigger}{46,354\xspace} 
\newcommand{\totalRecipesTrigger}{57,632\xspace} 
\newcommand{\coveredRecipeShares}{2,389,181\xspace} 
\newcommand{\totalRecipeShares}{2,760,341\xspace} 
\newcommand{\avgTriggersPerChannel}{3.5\xspace} 
\newcommand{\avgActionsPerChannel}{2\xspace} 
\newcommand{\medianTriggersPerChannel}{2} 
\newcommand{\medianActionsPerChannel}{1} 
\newcommand{\numDWORAttack}{4\xspace} 
\newcommand{\numDAttack}{22\xspace} 

\newcommand{\percentOverprivdChannels}{75\%\xspace} 

%% file: abstract2.tex
\begin{abstract}
Trigger-Action platforms are an emerging class of web-based systems that enable users to create automation rules (or \textit{recipes}) of the form, ``If there is a smoke alarm, then turn off my oven.'' These platforms stitch together various online services including Internet of Things devices, social networks, and productivity tools by obtaining OAuth tokens on behalf of users. Unfortunately, these platforms also introduce a long-term security risk: If they are compromised, the attacker can misuse the OAuth tokens belonging to millions of users to arbitrarily manipulate their devices and data. In this work, we first quantify the risk users face in the context of If-This-Then-That (IFTTT), a popular trigger-action platform that interfaces with $\totalChannels$ online services and provides over $200,000$ recipes. We perform the first empirical analysis of the OAuth-based authorization model of IFTTT using semi-automated tools that we built to overcome the challenges of IFTTT's closed source nature and of online service API inconsistencies. We find that $\percentOverprivdChannels$ of IFTTT's channels, an abstraction of online services, use overprivileged OAuth tokens, increasing risks in the event of a compromise. Even if the OAuth tokens were to be privileged correctly, IFTTT's compromise will not prevent their misuse. Motivated by this empirical analysis, we design, build, and evaluate \longsys (\sys), the first trigger-action platform where users do \textit{not} have to give it highly-privileged access to their online services. \sys splits the logically monolithic IFTTT architecture into a cloud service that users do not trust, and a set of clients. A user only has to trust the client she installs. Our design pushes the notion of fine-grained OAuth tokens to its extreme and ensures that even if the cloud service is controlled by the attacker, it cannot misuse the OAuth tokens to invoke unauthorized actions. Using such fine-grained tokens would normally lead to a drastic increase in the number of OAuth permission prompts. \sys avoids this increase by introducing the concept of a Transfer Token (XToken) and combining it with recipe-specific tokens. Our evaluation establishes that \sys poses modest overhead: it adds less than $15ms$ of latency to recipe execution time, and reduces throughput by $2.5\%$.

\end{abstract}

%% file: intro2.tex

\section{Introduction}
\label{sec:intro}

Trigger-Action platforms are a class of web-based systems that stitch together several online services
to provide users the ability to set up automation rules. These platforms allow users to setup rules like,
``If I post a picture to Instagram, save the picture to my Dropbox account.'' The ease of use and functionality
of such platforms have made them increasingly popular, and several of them (\eg If-This-Then-That~\cite{iftttSite},
Zapier~\cite{zapier}, and Microsoft Flow~\cite{msflow}) are on the rise. Furthermore, with the rise in popularity of connected physical devices like smart locks and ovens, we observe that many trigger-action platforms
have started adding automation support for physical devices, making it possible for users to set up
rules like: ``If there is a smoke alarm, then turn off my oven''~\cite{smokerecipe}. These platforms have privileged access to a user's online services and physical devices. If they are compromised, then attackers can \textit{arbitrarily} manipulate data and devices belonging to a large number of users to cause damage.

We quantify the risk users face if a trigger-action platform is compromised by performing the first empirical analysis of IFTTT's authorization model---a cloud-based closed source system that provides a trigger-action abstraction for end-users in the form of {\em recipes}.\footnote{On Nov $2^{nd}$, IFTTT changed some of its naming conventions. \eg $Recipe\rightarrow Applet$, $ Channel \rightarrow Service$. These changes do not affect the functionality of IFTTT or our results. \url{https://ifttt.com/m/meet-the-new-ifttt} provides a full description of these changes.} Running our example recipe above requires IFTTT to integrate with the smoke alarm (Nest) and the oven and obtain authorization to access them on the user's behalf. It achieves this integration using its {\em channel} abstraction~\cite{iftttSite}, through which IFTTT gains privileged access to the user's accounts on online services that in turn provide access to the smoke alarm and the oven. IFTTT and online services decide the privilege using the popular OAuth protocol~\cite{oauth1,oauth2}, where IFTTT requests a certain amount of privilege using the \texttt{scope} parameter, and gains OAuth tokens if users grant privilege. Therefore, IFTTT contains OAuth tokens for all user online services in its logically monolithic architecture~\cite{ifttt-engg}. If IFTTT is compromised, then the attacker can misuse the tokens and \textit{arbitrarily} manipulate data and devices. Furthermore, incorrect OAuth scoping can lead to overprivilege---either IFTTT channels may request broad scopes or the online services may only offer coarse-grained scopes. In either case, if IFTTT is compromised, the highly-privileged OAuth tokens will only increase the damage that attackers can cause.


We choose to study IFTTT for several reasons. First, it is a popular platform, supporting $\totalChannels$ channels as of \dateofdata, and offers more than $200,000$ recipes, many created by end-users~\cite{urCHITrigger}. Second, its trigger-action programming abstraction is well-suited to many desirable home automation behaviors~\cite{urtrigger}, and it supports $\iotChannels$ cyber-physical channels. Third, IFTTT is representative of a larger class of trigger-action platforms. Other systems such as Zapier~\cite{zapier} and Microsoft Flow~\cite{msflow} share the same design principles. Therefore, the results and lessons we learn from our empirical study of IFTTT are broadly applicable to the security design of other systems in this class.

Performing this empirical analysis is challenging for several reasons. First, IFTTT is closed source---we cannot inspect source code or even binary code to determine what scopes the channels need or request. Second, many channels use a specific opaque \texttt{scope=ifttt} OAuth authorization URL parameter, defined by the online service for the channel, and that does not reveal the APIs that the channel can access. Third, there is no defined scope-to-API mapping in the OAuth specification, making it difficult to compute the set of online service APIs to which a scoped token gives access. Fourth, online services do not document their APIs in consistent ways and do not provide unit tests, making large scale testing difficult. We built a semi-automated measurement pipeline that overcame the above challenges to obtain tokens of the same privilege that IFTTT uses and then exhaustively tested online service APIs to compute a conservative lower bound on the overprivilege that IFTTT channels exhibit. Our results indicate that $\percentOverprivdChannels$ of the channels examined have access to more operations than they need to support their triggers and actions. 

The presence of such tokens makes IFTTT an attractive target for attackers and increases risks if it is compromised. For example, our empirical analysis shows that an attacker can reprogram Particle chips and delete Google Drive files with a single HTTP call (\xref{subsec:overprivresults}). Thus, users are taking a significant long-term risk in granting IFTTT highly-privileged access to their online data and devices.

We show that this risk can be avoided, while getting the benefits of a cloud-based trigger-action platform. We design, implement, and evaluate \longsys (\sys), the first decoupled trigger-action platform whose compromise does not permit an attacker to arbitrarily invoke functions in online services.
\sys achieves this property by introducing the notion of {\em recipe-specific tokens}. A recipe-specific token can only be used to execute the specified recipe. If a compromised cloud service attempts to use a token to invoke an action without a valid trigger, the online service will deny that request. Thus \sys ensures that the cloud service: (1) can only invoke actions and triggers needed for the recipes it is executing; (2) can invoke actions only if it can prove to an action service that the corresponding trigger occurred in the past within a reasonable amount of time; and (3) cannot tamper with any trigger data passing through it undetected. 


Unless carefully designed, recipe-specific tokens can lead to an increased number of OAuth permission prompts because users would have to login and approve an OAuth scope request every time they create a recipe. The challenge is to gain the security of recipe-specific tokens but maintain the current IFTTT experience where users login to channels only once during a setup phase. \sys overcomes this challenge by introducing the concept of {\em transfer tokens} (XTokens). A client uses an XToken to automatically obtain a recipe-specific token, which it transmits to the cloud service for recipe execution.


\sys requires the untrusted cloud service to prove to the invoked action service that a trigger has occurred within a reasonable amount of time in the past. As the cloud service can be compromised, a possible design is to have the trigger service communicate directly with the action service to verify the occurrence of a triggering event. However, this introduces an undesirable dependency between the action and trigger services. \sys avoids that by using a lightweight cryptographic signature-based extension to the OAuth 2.0 protocol.

\noindent\textbf{Our Contributions:}
\begin{itemize}
\item We perform an empirical analysis of IFTTT's authorization model to quantify the risk that users face in the event that it is compromised:

\begin{itemize}
\item  Based on an analysis of authorization sessions of $\scrShotChannels$ online services that IFTTT integrates with, we characterize: (1) the descriptiveness of OAuth scope requests (\xref{sec:bg}, \xref{sec:empirical}), and (2) the level of control these services provide to users when IFTTT requests scopes. Our analysis reveals that most online services  ($101/\scrShotChannels$) provide a good or acceptable explanation of the scopes being requested. Unfortunately, for many online services ($\binaryOAuth/\scrShotChannels$), a user is only given an all-or-nothing choice of accepting all scope requests or none at all.  Therefore, for most online services, even though users are told what level of access is being requested, they are not provided the means to restrict the amount of privilege they grant to IFTTT. 


\item We find that $107/\scrShotChannels$ channels use an opaque scope that the online service provides, such as \texttt{generic, null}, or \texttt{ifttt}. This can lead to overprivilege. Therefore, we perform an in-depth analysis
of the overprivilege of all externally measurable IFTTT channels  ($\measurableChannelsAll$ of them) and obtain a conservative lower bound.
We study all $\numApisStudied$ APIs across $\numChannelsStudiedAll$ channels, including $\numChannelsStudiedIoT$ higher-risk cyber-physical channels, and find that \overprivdChannelsAll channels including
10 cyber-physical channels have access to APIs that they do \textit{not} need
to implement their functionality. Our analysis covers $\recipeCoverageTrigger$
of all recipes involved in the set of
$\measurableChannelsAll$ channels (\xref{subsec:overprivresults}). Examples of overprivileged channels
include well-known services like Facebook, Twitter, and Google Drive and cyber-physical services like Particle, and MyFox Home Control. Using such overprivileged access, a potentially compromised IFTTT platform can reprogram a Particle chip's firmware or delete files on Google Drive arbitrarily with a single single HTTP call.
\end{itemize}

\item Motivated by this empirical analysis, we designed and implemented \longsys, the first decoupled trigger-action platform where users do not have to trust the platform with highly-privileged access to their online services (\xref{sec:defense}). \sys splits the logically monolithic IFTTT architecture into an untrusted cloud service that executes recipes at scale, and a set of clients that help users create recipes in a secure manner. \sys is based on cryptographic extensions to the OAuth protocol that only allow the cloud service to execute user recipes, even if it is attacker-controlled. Our design introduces techniques that enable trigger-action functionality without putting users at risk of attackers arbitrarily manipulating data and devices. Recipe-specific tokens also reduce the platform's liability in the event of a compromise.

\begin{itemize}
\item Our evaluation of \sys shows that performance overhead is modest (\xref{sec:eval}): Each recipe requires less than $3.5KB$ additional storage space and imposes less than $7.5KB$ of transmission overhead per execution. 
\item \sys adds less than $15ms$ of latency to recipe execution time. For recipes on IFTTT, which typically send emails, SMSs, or invoke actions on physical devices or on online services over a network, we consider this additional latency to be acceptable. \sys reduces throughput by $2.5\%$ for recipe execution. 
\end{itemize}

\end{itemize}

Although \sys requires online services to support recipe-specific tokens and to use cryptographic extensions to the OAuth 2.0 protocol, there are several options to help developers make the transition. One option that we followed in our implementation is to design a library that only requires adding a single annotation above HTTP methods in the server (\xref{sec:eval}). Another option is to construct a trusted proxy for the online services (\xref{sec:discussion}). Furthermore, as these platforms are still emerging, our design gives system implementers an opportunity to construct secure trigger-action platforms from the ground up, and represents an important first step towards securing trigger-action platforms.

%% file: bg.tex
\section{If-This-Then-That}
\label{sec:bg}

\begin{figure}[!tb]
	\center
	\includegraphics[width=\columnwidth]{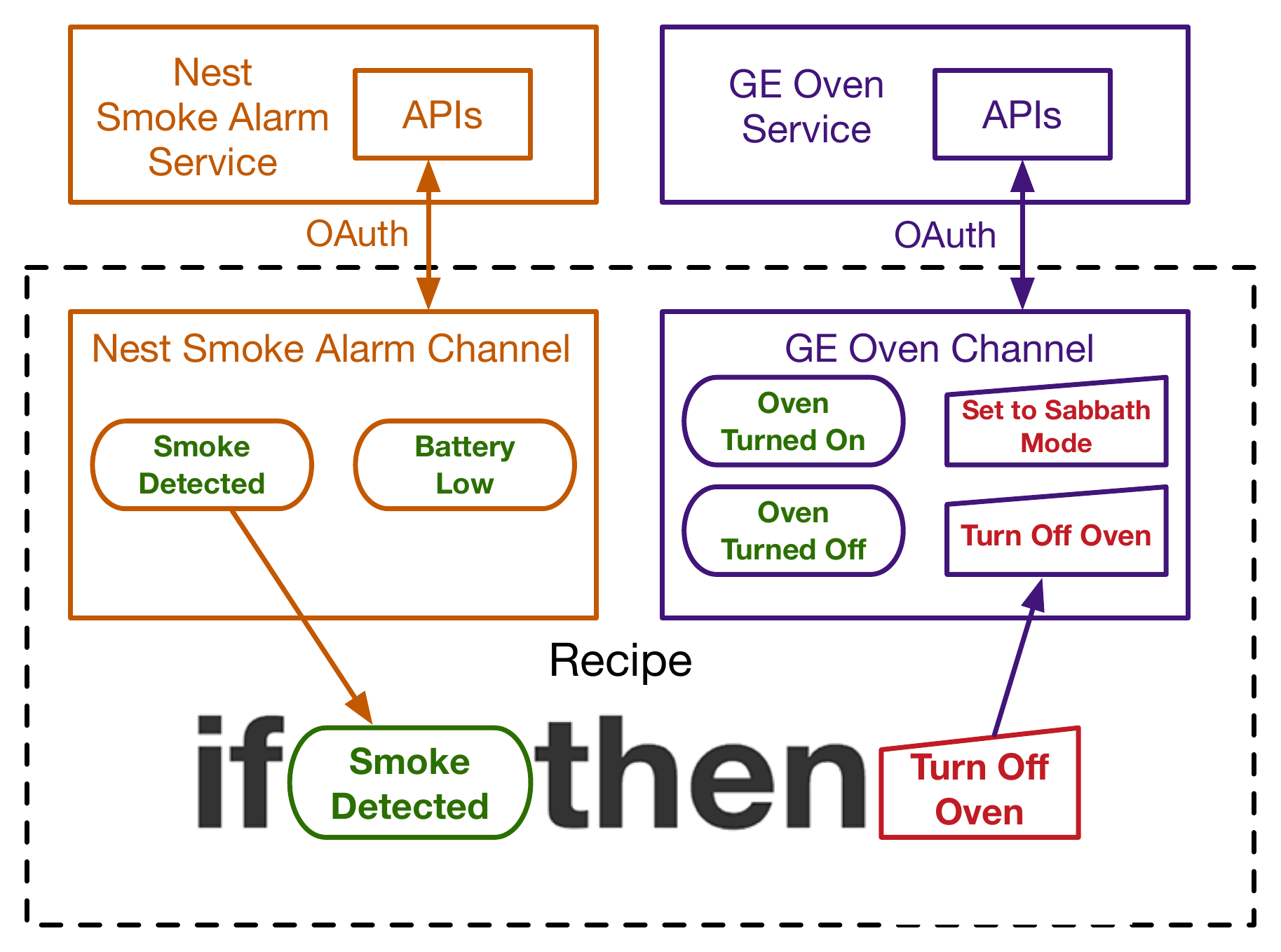}
	\caption{An overview of IFTTT architecture in the context of a recipe. Online services have a channel inside IFTTT. These channels gain access to online service APIs by acquiring an OAuth token during the channel connection step. A recipe combines a trigger and an action.}
	\label{fig:ifttt_arch}
\end{figure}

IFTTT is a popular trigger-action platform that makes it easy for end-users to set up interactions
between various online services and IoT devices to achieve useful automation. 
IFTTT works by communicating with the REST APIs that these services expose.
Figure~\ref{fig:ifttt_arch} shows the high-level IFTTT architecture. 
It has the following four architectural components:

\begin{itemize}
\item \textbf{Channel:} A channel represents part of an online service's set of APIs on the IFTTT platform. Users connect
channels to their IFTTT accounts---a process that involves user authorization.
For example, a user with a Facebook account must authorize the IFTTT Facebook 
channel to communicate with the corresponding Facebook account. Channels communicate with online services using
REST (Representational State Transfer) APIs operating over HTTP(S). These online services 
use the popular OAuth protocol to enforce authorization~\cite{oauth1,oauth2}.
Users must connect several such channels, before they can accomplish
any useful work. Either IFTTT developers or service providers can implement channels. In the latter case, IFTTT exposes a separate API
to channel writers to help them integrate their online service with IFTTT.\footnote{This API is currently in private beta~\cite{iftttpartners}.}

\item \textbf{Trigger:} A channel may provide triggers, which are events that occur in the associated online service. ``A file was uploaded
to a cloud drive'' or ``smoke alarm is on'' are examples of triggers.

\item \textbf{Action:} A channel may also provide actions. An action is a function (or set of functions) that exists in the API of the online service.
Examples of actions include ``turning on or off a connected oven'' or ``sending an SMS.''

\item \textbf{Recipe:} Recipes are at the core of the IFTTT user experience, and they are the core functionality that IFTTT enables.
A recipe stitches together two channels to achieve useful automation. It has two pieces. The ``If'' piece represents a trigger
or an event that occurs on an online service. The ``Then'' piece represents an action that should be executed on the online service.
For example, ``If there is a smoke alarm, then turn off my oven.'' This recipe integrates the smoke alarm channel's ``alarm is on''
trigger with the oven channel's ``turn off the oven'' action. Although IFTTT only permits a single trigger and a single action
in a recipe, other trigger-action platforms offer multiple triggers and actions in the same recipe.
\end{itemize}

There is a one-to-one correspondence between online services and IFTTT channels.
For example, the Google Drive online service corresponds to the Google Drive IFTTT channel.
IFTTT has $\totalChannels$ channels as of \dateofdata. 


\begin{figure}[!tb]
	\center
	\includegraphics[width=\columnwidth]{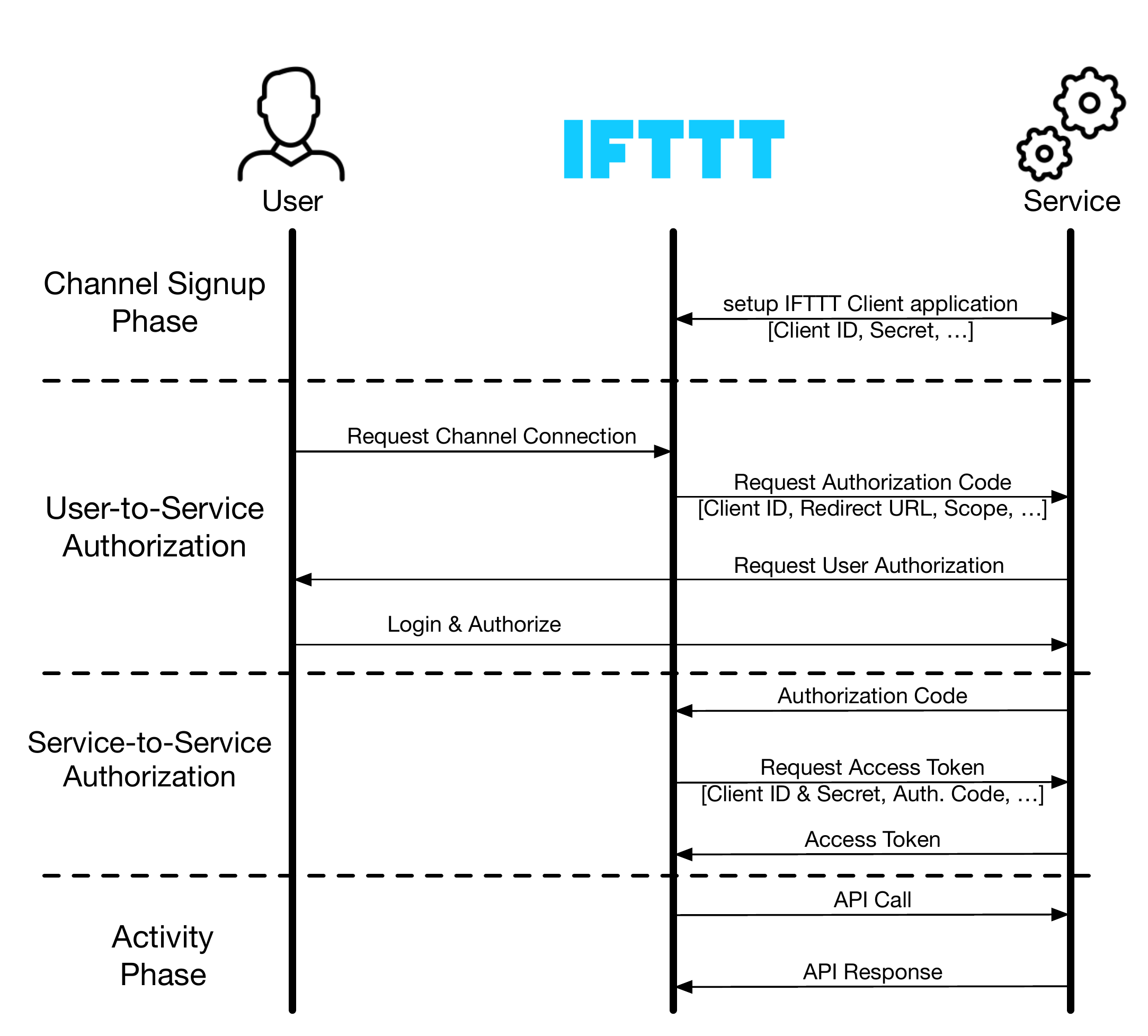}
	\caption{IFTTT's authorization model has four phases. Channel developers create client applications for the online service that results
    in the online service assigning a client ID and secret to the application. Then, IFTTT
    initiates an authorization workflow. The OAuth 2.0 authorization code flow is a popular choice, and it results in IFTTT gaining a 
      scoped bearer token that authorizes a channel to invoke APIs on 
    an online service. Users are prompted to approve or deny scope requests during this process.}
	\label{fig:oauth}
\end{figure}

\noindent \textbf{Authorization Model.} Online services protect their REST APIs using authorization protocols. OAuth is a popular
choice that enables an online service to provide third parties with secure delegated access to its APIs. IFTTT must obtain authorization to communicate with online services that its channels
represent; and therefore it must follow the OAuth authorization workflow. Figure~\ref{fig:oauth} shows the IFTTT authorization model.
It has four steps. 

First, a channel developer (IFTTT or the
online service provider itself) must create a client application for the online service's REST API. This client application
represents an IFTTT channel on the online service. During the sign-up phase, the online service assigns a client ID and a secret
that IFTTT uses during the authorization workflow.

Second, a user initiates a channel connection within the IFTTT administrative interface and this causes IFTTT to initiate the
OAuth 2.0 authorization code flow---the recommended workflow for server-to-server authorization---that results in
IFTTT requesting the corresponding online service for a short authorization code on behalf of the user. 
IFTTT passes a client identifier value, a redirect URI, and a scope
value as part of the HTTP(S) request. The scope value represents the level of access IFTTT is requesting to operate a channel. This authorization request results in the user being presented with an OAuth permissions screen that explains the scope that IFTTT is requesting. As the OAuth protocol does not specify the design of the permissions screen, the screen design, scope explanations, and UI options to modify the requested scopes is at the discretion of the online service.

Third, assuming the user accepts the scope request, the online service redirects to the IFTTT-provided redirect
URI with a short authorization code as an argument. Fourth, IFTTT exchanges the authorization code, client ID, and client secret for an access token using server-to-server communication. Finally, IFTTT can then use the OAuth bearer token to initiate API calls on the online service to implement channel functions.

Although OAuth 2.0 is by far the most popular authorization protocol that IFTTT uses, there are online services that use
OAuth 1.0a. This protocol does not have explicit scoping as part of the authorization workflow, but a similar concept is available
when a client application signs up for the online service's API. During the client application sign-up phase, the developer can
choose scopes to enable. For example, Twitter uses OAuth 1.0a, and it provides a settings item that allows a developer to change
the access level of the client application, and hence, change the scope of any tokens issued in the future. 


\subsection{Potential for Overprivilege}
Ideally, IFTTT should be authorized with only enough privilege to run a given user's set of recipes.
Any access rights beyond what it requires to run a user's recipes is overprivileged access. 
Based on the authorization model and IFTTT architecture discussion above, we observe the potential for two kinds of systemic overprivilege that
stems from IFTTT's design. We discuss them below.

\noindent{\textbf{Recipe-Channel Overprivilege.}} An IFTTT recipe only requires a single trigger and a single action. However, one channel can support multiple triggers
and actions. Table~\ref{tab:particledrivetrigacts} shows the set of triggers and actions for two example channels---Particle
and Google Drive. Therefore, if a user adds a recipe that uses a subset of triggers
and actions of the associated channels, all additional triggers and actions that the channel implements is overprivileged access.
That is, IFTTT acquires the right to invoke online service APIs that it does not need for executing the recipes of a given user. 
This kind of overprivilege stems from the IFTTT
design choice that channel authorization is not recipe-specific.

\begin{table}[!tb]
\center
\begin{tabular}{c C{3cm} C{3cm}}
\toprule
\textbf{Channel} & \textbf{Triggers} & \textbf{Actions} \\ \toprule
\multirow{4}{*}{Particle} & New event published. & \multirow{2}{*}{Publish an event.} \\
										 & Monitor a variable.     & \\
										 & Monitor a function result. & \multirow{2}{*}{Call a function.} \\ 
										 & Monitor your device status. & \\
\midrule
\multirow{4}{*}{Google Drive}  & \multirow{4}{*}{NONE}  & Upload file from UR.L \\
 & & Create a document. \\
 & & Append to a document. \\
 & & Add row to spreadsheet \\
\bottomrule
\end{tabular}
\caption{Triggers and Actions for the Particle and Google Drive Channels.}
\label{tab:particledrivetrigacts}
\end{table}

\noindent{\textbf{Channel-Online-Service Overprivilege.}} The second type of overprivilege we observe is related to the authorization between IFTTT and an online service.
As discussed, IFTTT provides channels that have trigger and action functionality. Furthermore, IFTTT must
request authorization to a set of online service APIs to implement a channel's functionality. Ideally, the IFTTT channel
should only have the authorization needed to implement its functionality. However, in practice, incorrect scoping 
can lead to IFTTT gaining authorization to access APIs it does \textit{not} need to implement its trigger and action functionality.
This can occur if the online service does not provide granular scopes, forcing IFTTT to request coarse-grained overprivileged
access. This can also occur if IFTTT incorrectly requests overprivileged access even if the online service provides fine-grained
scopes.


\subsection{Examples of Overprivilege}
\label{subsec:overprivexamples}
Here we show the potential for the two types of overprivilege discussed above using two case studies.

\begin{figure}[!tb]
\center
\includegraphics[height=0.5\columnwidth]{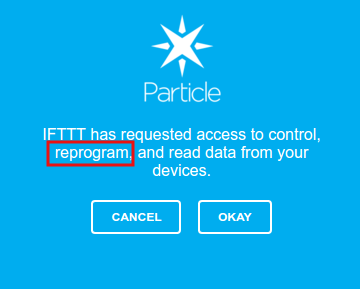}
\caption{Particle OAuth permissions prompt. This indicates that the IFTTT Particle channel will have the ability to reprogram
a Particle chip even when there are no triggers or actions that support such functionality, leading to overprivileged access for
the channel.}
\label{fig:particle}
\end{figure}

\noindent \textbf{Particle.} This is a DIY electronics platform that offers small form-factor chips with a microcontroller, memory,
and an Internet connection. Particle supports writing custom firmware for its chips and exposes a REST API to make
the chips remotely accessible. IFTTT supports the Particle channel. Table~\ref{tab:particledrivetrigacts} shows the set of triggers
and actions for this channel. If a user's recipes do not use all the triggers and actions, then that channel exhibits recipe-channel overprivilege.
Figure~\ref{fig:particle} shows the OAuth permissions screen a user sees while connecting the Particle channel to IFTTT. Clearly, the channel is requesting privilege to ``reprogram'' a Particle device. However, based on the triggers and actions from Table~\ref{tab:particledrivetrigacts}, the channel does not offer any such operation in recipes, and hence exhibits channel-online-service overprivilege. If IFTTT is compromised, an attacker can use this level of access to reprogram a chip even though there are no channel operations and no recipes that offer such functionality.

\begin{figure}[!tb]
\center
\includegraphics[width=\columnwidth]{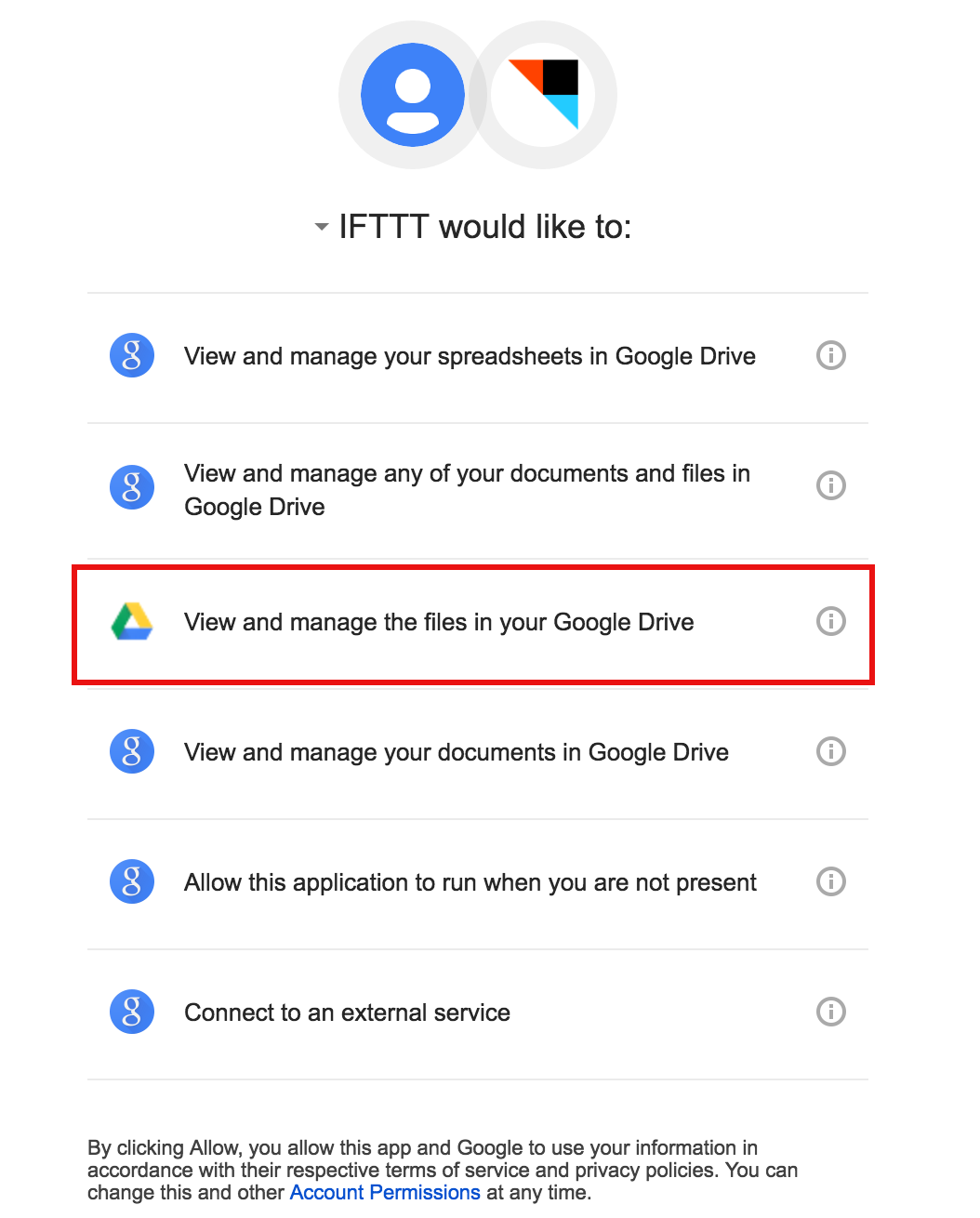}
\caption{Google Drive OAuth permissions prompt. ``View and manage the files in your Google Drive'' implies the ability to
``Upload, download, update, and delete files in your Google Drive'' as per Google Drive API documentation. This is overprivileged
access since no triggers and actions of the channel allow deleting files.}
\label{fig:googledrive}
\end{figure}

\noindent \textbf{Google Drive.} This is a well-known online service that offers cloud storage for various types of files.
Similar to Particle, if a user's Google Drive recipes do not use all the triggers and actions of the corresponding IFTTT channel (see Table~\ref{tab:particledrivetrigacts}), then that channel exhibits recipe-channel overprivilege. Figure~\ref{fig:googledrive} shows the OAuth permissions screen a user sees while connecting the Google Drive channel to IFTTT. The prompt indicates that IFTTT will be able to ``View and manage the files in your Google Drive.'' Based on Google Drive API documentation, this implies the ability to ``Upload, download, update, and delete files in your Google Drive.'' However, the Google Drive channel does not offer any triggers or actions that delete files. Thus we conclude that the IFTTT Google Drive channel exhibits channel-online-service overprivilege.


Therefore, based on these case studies, we observe the potential for overprivilege in terms of recipes operating with channels and in
terms of channels interfacing with the online service APIs. We focus our empirical overprivilege analysis on these two types of
overprivilege.

%% file: empirical.tex
\section{Empirical Overprivilege Analysis of IFTTT}
\label{sec:empirical}
We studied $\totalChannels$ channels, covering $\totalRecipes$ recipes and performed an in-depth
overprivilege computation for $\numChannelsStudiedAll$ channels, including $\numChannelsStudiedIoT$ cyber-physical channels, achieving
a coverage of $\recipeCoverageTrigger$ of all recipes involved in a set of $\measurableChannelsAll$ channels that can be studied in-depth.
Our findings are two-fold. First, based on an analysis of the triggers and actions supported by $\totalChannels$ channels, we found
that on average, each channel has $\avgTriggersPerChannel$ triggers and $\avgActionsPerChannel$ actions. As a recipe only
needs one trigger and one action, if a user's set of recipes do not use all channel triggers/actions, then there is overprivilege.
Second, we find that $\overprivdChannelsAll$ (out of $\numChannelsStudiedAll$) channels have access to online service APIs that
they do not need to implement their trigger/action functionality. We present details on our dataset and how we constructed it, and we present
our semi-automated overprivilege measurement pipeline, along with details on its output.

\subsection{Dataset and Measurement Setup}
\label{subsec:dataset}


There are two parts to our dataset: recipes and channels. We used the recipe dataset from Ur \etal~\cite{urCHITrigger} that
contains \totalRecipes recipes (after filtering out recipes that refer to defunct channels), and we created our own
dataset of \totalChannels channels that consists of trigger and action details. We created this dataset of channels on \dateofdata. 
We also scraped the IFTTT website to download trigger and action details (description, input arguments) for all channels of the dataset.

Subsequently, we created a test IFTTT account and an account for each online service. We then manually connected each channel by following 
the OAuth workflow. We were able to connect \connectedChannels channels to our test IFTTT account, out of which, we recorded 
authorization sessions for \scrShotChannels channels.
An authorization session consists of two items: (1) The OAuth authorization URL that IFTTT uses to initiate authorization with
the online service to receive a token, and (2) A screenshot of the OAuth permissions prompt.
We were unable to connect or record authorization sessions for \numChannelsCouldNotAnalyze channels due to the following reasons:
\begin{itemize}
\item \textbf{Channels that are connected by default or do not require authorization:} Channels such as {\em SMS}, {\em weather}, and {\em ESPN} 
that are not tied to any 3rd party user account fall in this category. We 
observed $\defaultConnected$ such channels among $\totalChannels$ in our study. We exclude these channels from further analysis.
	
\item \textbf{Channels that require specific physical devices for connection or are behind a pay-wall:} Channels such as \emph{Feedly} and \emph{D-Link Siren}
that require either purchasing a physical device or purchasing a subscription to complete the connection to IFTTT fall in this category. 
We observed $\paywallChannels$ such channels. To minimize the cost of our research project, we exclude them from further analysis.
We note that we include channels with physical devices in our analysis if they could be connected to our test IFTTT account without a physical
device being present. We also include the Alexa and SmartThings channels for which we purchased physical devices.

\item \textbf{Channels that are mobile only and communicate only through IFTTT mobile app:} Channels such as \emph{iOS Contacts} and \emph{Android Location} that only
operate on a mobile platform such as Android and iOS fall in this category. We observed $\mobileChannels$ such channels. 
These channels rely on the permission architecture of the underlying mobile platform and do not use OAuth. Therefore, we exclude
them from further analysis.
	
\item \textbf{Defunct or Malfunctioning Channels:} Channels such as \emph{Home8} and \emph{iSmartAlarm} that were either defunct or malfunctioning during
the connection process fall in this category. We observe $\defunctChannels$ such channels, which we exclude from further analysis.

\end{itemize}

Table~\ref{fig:appBreakdown} presents a breakdown of the connection status of our $\totalChannels$ channels. 
We focus the rest of our analysis on the $\scrShotChannels$ authorized channels, and their associated recipes. 

\begin{table}[tb!]
	\center
	\begin{tabular}{c c C{.7cm} C{1cm}  C{.5cm}  c c}
		\toprule
		\textbf{Status} & \textbf{Authorized} &	\textbf{Req. Device} & \textbf{Mobile App Only} & \textbf{No Auth.} & \textbf{Pay-Wall} & \textbf{Other} \\
		\toprule
		\textbf{Channels} & $\scrShotChannels$ & 57 & 33 & 40 & 10 & 29\\ \bottomrule
	\end{tabular}
		\caption{Channel connection status. We were able to record authorization session information for $43\%$ of all channels in our dataset.
      $19\%$ of channels require a physical device to be present during connection, $11\%$ gain authorization through a mobile app's native
        permission model, $19\%$ of channels connect without requiring authorization, $3\%$ sit behind a pay-wall and could not be connected,
    and $10\%$ were either defunct or malfunctioning at the time of our analysis.}
		\label{fig:appBreakdown}
\end{table}

\subsection{Initial Observations}
\label{subsec:initobserve}
We recorded authorization sessions (screenshot of OAuth permissions prompt and authorization URL) for $\scrShotChannels$ channels connected to 
our test IFTTT account. Based on this data, we make initial observations about: (1) Permission prompt descriptiveness 
and user control, and (2) OAuth token scopes.

\begin{itemize}
\item OAuth permission prompts serve as potential control points for users to enforce fine-grained control over the privilege that third parties gain. We analyzed whether the permission prompt screens in our dataset offer fine-grained control by manually interacting with the UI and checking for options to modify the requested privilege. Out of $\scrShotChannels$ channels we studied, the online services for $\binaryOAuth$ of those channels provide the user with an all-or-nothing choice (authorizing the channel or not), even though $\binarySingleTrigAct$ of them have multiple triggers or actions. The online services of $\timedOAuth$ channels (Evernote and LinkedIn) allow the user to select the time duration for the authorization, and only $\fineOAuth$ online services (AT\&T M2X, Facebook, Pushover, and SmartThings) provide fine-grained control over data or devices that were being shared with IFTTT.

\item OAuth permission prompts are an opportunity for online services to explain the privilege that third parties are requesting. We analyzed whether online services in our dataset provide an adequate explanation of privilege by comparing the description of the channels triggers and actions with text in the prompt. Out of $\scrShotChannels$ channels, only $\goodinfo$ of the corresponding online services provide an adequate description of data that was being shared; $\okinfo$ online services provide some description that is either vague or insufficient in explaining their function to the user; and $\noinfo$ provide no information on data that was being shared with the user.
\end{itemize}

Based on these measurements, we conclude that even though online services provide some description of the privilege that third parties like IFTTT request, they do not offer the user control over those requests. This forces users to entrust IFTTT with highly-privileged access to their devices and data, thus increasing the risk they face if attackers compromise IFTTT.

Next, we computed the average and median number of triggers and actions for $\totalChannels$ channels. We find that channels have on average $\avgTriggersPerChannel$ triggers per channel (median = $\medianTriggersPerChannel$), and
an average of $\avgActionsPerChannel$ actions per channel (median = $\medianActionsPerChannel$) with a long-tail distribution (See Appendix~B for a graph). As an IFTTT recipe only uses a single trigger and a single
action, most channels exhibit recipe-channel overprivilege. However, if users create recipes that involve all triggers and actions of all connected channels, then this type of overprivilege does not exist any more.


Finally, we determined the distribution of various OAuth protocol variants in use. Out the set of $\scrShotChannels$ connected and authorized channels, $113$ online services that correspond to these channels use OAuth 2.0, making it the most popular variant (the remaining $8$ use OAuth 1.0 and its variants, and we could not determine the protocol being used for $7$ channels due to lack of information in the authorization sessions). Therefore, we drilled down and analyzed the OAuth 2.0 scopes being requested because they define the privilege an IFTTT channel requests from the online service (\xref{sec:bg}).

Table~\ref{tab:scopebreakdown} shows the breakdown of these scopes. Similar to what we observe based on OAuth permissions prompts, we see that $107$ channels request a scope that is generic, null, or simply specified as ``ifttt''---clearly coarse-grained privilege requests. Furthermore, due to IFTTT architecture, fault cannot clearly be placed on IFTTT or on the online services since a channel may be written by either IFTTT developers or the online service provider. If an online service provider only has coarse-grained tokens, then a channel only has the option to request overprivileged access, irrespective of whether IFTTT developers or the service provider creates the channel. We also observe that only $21$ online services out of $\scrShotChannels$ offer fine-grained scoping. Table~\ref{tab:scopeExample} shows examples of fine-grained scopes for channels in our dataset.

\begin{table}[tb!]
	\center
	\begin{tabular}{c c c c c}
		\toprule
		\textbf{Scope} & \texttt{ifttt} &	\texttt{null} & \textbf{Generic} & \textbf{Fine-Grained} \\
		\toprule
		\textbf{Channels} & 77 & 26 & 4 & 21\\ \bottomrule
	\end{tabular}
	\caption{$83\%$ of online services do not provide fine-grained scoping and only provide opaque scopes like generic, null, or ifttt. We show examples
  of fine-grained scopes in Table~\ref{tab:scopeExample}. Examples of generic scopes are ``spark'' or ``app.''}
	\label{tab:scopebreakdown}
\end{table}

\begin{table}[!tb]
	\centering
	\begin{tabular}{c C{1.5cm} C{4cm}}
    \toprule
		\textbf{Channel} & \textbf{Allows User Control} & \textbf{Scope}  \\ \toprule
 AT\&T M2X		&   \cmark    &       IDENTITY, GET, POST \\\midrule
Facebook		&    \cmark    &     manage\_notifications, manage\_pages, public\_profile, publish\_actions, user\_about\_me, user\_activities, user\_events, user\_friends, user\_location, user\_photos, user\_posts, user\_status, user\_website ٓ\\ \midrule
Netatmo Welcome & \xmark & read\_camera, write\_camera \\ \midrule
Pinterest & \xmark & read\_public, write\_public, read\_secret, write\_secret, read\_relationships, write\_relationships     \\ \midrule
Youtube & \xmark & https://www.googleapis.com/auth/youtube \\
  \bottomrule
	\end{tabular}
	\caption{Examples of fine-grained scopes requested by channels. $21$ channels in total use fine-grained scopes when requesting tokens. Only $4$ online services that correspond to these channels allow the user to exert control over them.}
	\label{tab:scopeExample}
\end{table}

A significant fraction of the channels we study use fairly coarse-grained, opaque scopes. Therefore, we drilled down further to determine the privilege these scopes represent, and obtain a conservative lower bound on the overprivilege a channel exhibits
over a given online service in terms of the APIs the online service exposes (channel-online-service overprivilege). This involved examining in detail the set of APIs a service provides. There are several challenges while performing this analysis. We discuss them next.

\input{measurechanneloverpriv}
\input{overprivresults}

%% file: measurechanneloverpriv.tex
\subsection{Measuring Channel-Online-Service Overprivilege}
\label{subsec:measurechanneloverpriv}
The scope of the OAuth tokens IFTTT gains to operate a channel
defines the privilege that channel has over the corresponding online service.
As we discuss in \xref{sec:bg}, a channel implements various triggers and actions. Each of the triggers
and actions will correspond to a set of one or more online service APIs. For example, the Particle
channel~\cite{particlechannel} from the DIY electronics category provides triggers to monitor device status and function results. It also provides actions to call functions and to publish events. Ideally, the channel would have privilege to only
perform these triggers and actions. However, due to the opaque nature of the scopes (see \xref{sec:empirical}), it is 
difficult to understand whether this is indeed the case. 

Therefore, we measure the \textit{Channel-Online-Service} overprivilege
on IFTTT. A channel exhibits such overprivilege if it can access APIs that are \textit{not} needed to implement
its triggers and actions.
Given the opaque nature of the scopes, it is not feasible to
determine what APIs the channel can access on the online service by simply reading the documentation. Furthermore, many services
do not document the scope-to-API mapping. IFTTT is also closed source, and most online services
are closed source too, ruling out any source-code inspection to determine the access control policy for a given scope.

Our measurement strategy is to capture scoped tokens identical to what IFTTT uses for a channel and then use those tokens
to exhaustively test the online service APIs to compute the set of APIs the tokens give access to. We encountered
the following challenges while performing this measurement:

\begin{itemize}
\item IFTTT obtains the OAuth token for a channel using server-to-server communication. This precludes the possibility
of intercepting the tokens that IFTTT itself uses. We verified that neither the IFTTT client side code nor the IFTTT
mobile app is directly involved in the OAuth authorization workflow---all authorization occurs on IFTTT's back-end cloud
service.

\item Online services do not document their APIs in consistent ways, making automated large scale testing difficult.

\item The types of online service APIs are very diverse, making it challenging to manually create input arguments. 
Furthermore, online services generally do not provide any unit tests with appropriate testing data, making it difficult
to distinguish input argument errors from permission errors.

\item Many online services have proprietary APIs that are only exposed to IFTTT. There is no public information on these APIs
precluding any kind of measurement.
\end{itemize}

We built a semi-automated measurement pipeline that addresses all challenges to an extent, except the last one because the online
service did not provide any documentation.

\begin{figure*}[t!]
\center
\includegraphics[width=0.9\textwidth]{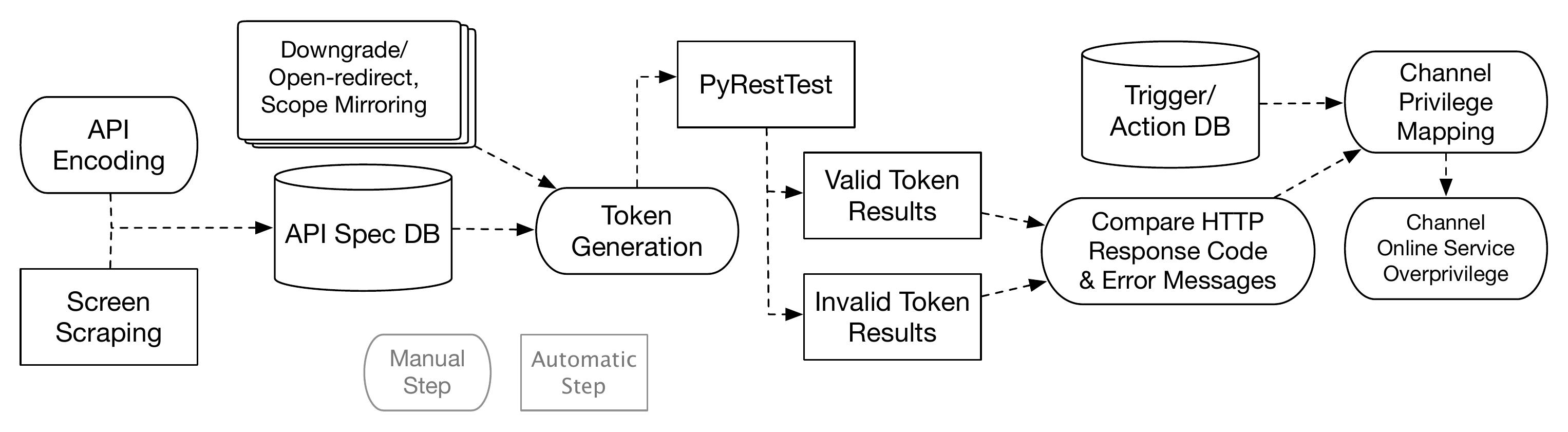}
\caption{Our semi-automated measurement pipeline to compute channel-online-service overprivilege. We use valid and invalid tokens
to distinguish input argument errors from authorization errors. }
\label{fig:measurearch}
\end{figure*}

Figure~\ref{fig:measurearch} shows our measurement tool's architecture. It takes as input an API specification database. We create
this database using two techniques: (1) Manual encoding of a REST API into our database format and (2) Automated screen-scraping
of the online service's API documentation if it is sufficiently regular HTML. The automated step helps us partly overcome the challenge
of online services not documenting their APIs in consistent ways.

The token generation step reads in the API specification
database and then creates a testing specification that includes two types of tokens: (1) A valid token for the online service and 
(2) An invalid token for the online service, mutated from the valid token. Then the tool uses PyRestTest~\cite{pyresttest}, an automated REST API 
testing tool to execute the test specification. We use these two types of tokens to address the challenge
of distinguishing input argument errors from permission errors. Since there is a large number of APIs and a large number
of online services, it is not possible to scalably create valid input arguments for a given API. The only scalable and automated
option is to create randomized inputs. However, this leads to API errors because these randomized arguments are often not
what the API expects to function correctly. Therefore, we needed a reliable way to distinguish input validation errors from
permission errors. Furthermore, since the OAuth
standard does not mandate specific error codes for specific conditions, it is often an implementation-dependent choice on
what HTTP error code to use. Therefore, our pipeline performs testing with a known invalid token and with a known valid
token. Then we manually analyze the change in error code and error message payload to decide whether the given valid token has permission
to access the API function in question or not.

Token generation also overcomes the challenge of server-to-server IFTTT authorization using four techniques:

\begin{itemize}
\item \textbf{Scope Mirroring:} A lot of channels use opaque scopes like ``ifttt'' (Table~\ref{tab:scopebreakdown}). Our
strategy to analyze opaque scopes is to create a mirror IFTTT-like application for a particular online service and then request 
the same opaque scope. This involves
following the OAuth authorization workflow manually which yields an identically scoped token that IFTTT itself uses for that online service.
In the case where the online services do not have opaque scopes, we observe that the scopes that IFTTT
uses can be expressed using the publicly available scopes. Therefore, our strategy is to make our IFTTT mirror application request
the same public scopes that IFTTT requests and to verify that the resulting OAuth permission screen looks identical to the one
we captured as part of the authorization session in the data collection step (\xref{subsec:dataset})

\item \textbf{Downgrade with Open Redirect:} The OAuth 2.0 protocol supports two major authorization workflows: authorization code grant and
implicit grant. The authorization code grant is the more secure option since it requires a client secret that is never revealed
to client-side code in IFTTT's case. However, we observe that many OAuth implementations also support the implicit grant flow---no client
secret is required, and a third party can obtain a token simply with a redirect from the authorization server. Furthermore, the implicit
grant is vulnerable to the open redirector attack~\cite{oauth2}, where an attacker can replace the \texttt{redirect\_uri} component
of the authorization URL with an attacker controlled domain. Therefore, our strategy is to attack our own test IFTTT accounts on the
online service by pretending to be an IFTTT-like application and performing a downgrade attack with open redirectors. 
The result is that we obtain an identically scoped token to what IFTTT uses. For example, consider the Ubi channel authorization
URL: 

\texttt{\url{http://portal.theubi.com/oauth/authorize?client_id=REDACTED&redirect_uri=https://ifttt.com/channels/ubi/authorize&response_type=code&scope=ifttt&state=VALUE}} 

A downgrade attack URL would be of the form: 

\texttt{\url{http://portal.theubi.com/oauth/authorize?client_id=REDACTED&redirect_uri=ATTACKER_URI&response_type=token&scope=ifttt&state=VALUE}}

Notice the change in the \texttt{redirect\_uri} parameter and the \texttt{response\_type} parameter.

\item \textbf{Downgrade only:} This strategy is similar to the above, except that the online service is vulnerable to a downgrade attack but 
not vulnerable to an open redirector attack. This prevents us from receiving the access token on an attacker-controlled domain.
Instead, we used a man-in-the-browser attack to read the OAuth token.

\end{itemize}

Once our pipeline produces two API testing results---one with a valid token and one with an invalid token, we manually
compare the outputs to produce a \textit{privilege mapping}. This mapping encodes whether the valid
token has access to a given API. The two API resting results make it trivial to observe the change in error code and error message
to determine whether the API failed due to a permission error or due to a scoping error.

Once we have a privilege mapping, we manually map channel operations to online service APIs.
This step involves examining the documentation of the online service APIs and
the trigger/action documentation, followed by \textit{conservatively} determining whether a particular API can be used to implement
a trigger or action. Our goal here is to obtain a conservative lower-bound on channel-online-service overprivilege. At the end of this step,
any API for which the token provides access, but is not needed to implement any trigger/action for the channel, is marked as
an overprivileged API.

%% file: overprivresults.tex
\subsection{Channel-Online-Service Overprivilege Results}
\label{subsec:overprivresults}

\begin{figure}[!tb]
	\center
	\includegraphics[width=1.05\columnwidth]{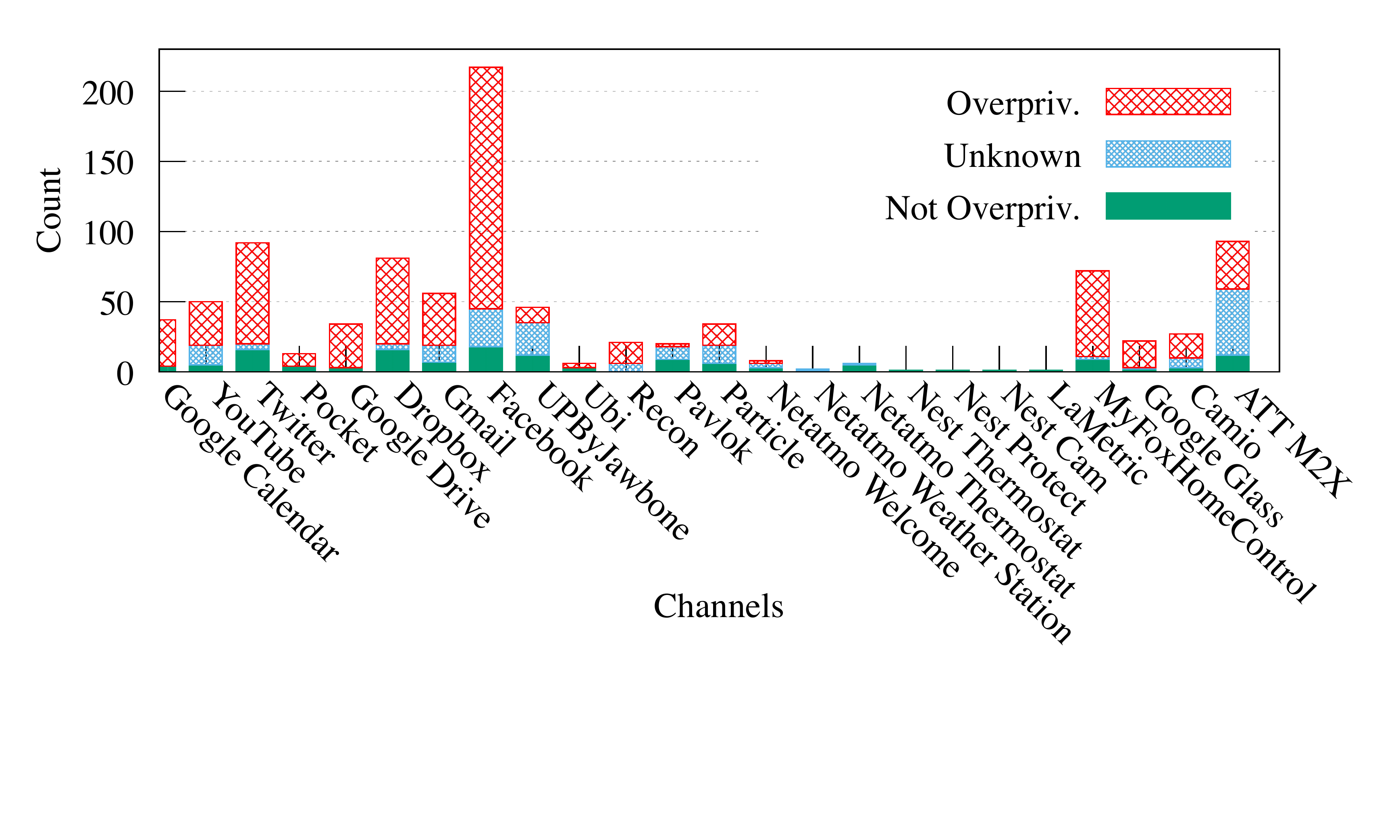}
	\vspace*{-15mm}
	\caption{Number of API functions accessible to IFTTT channels based on their privilege. $66.42\%$ of API functions accessible to IFTTT channels are not used in any trigger or action.}
	\label{fig:apiCount}
\end{figure}

To compute overprivilege in depth, we applied our measurement pipeline on $\numChannelsStudiedAll$ channels out of a total
of \measurableChannelsAll channels that can be measured. 
Figure~\ref{fig:apiCount} shows a breakdown of our channel-online-service overprivilege results. Our measurement tool outputs a privilege mapping
for \numApisStudied APIs across \numChannelsStudiedAll channels. We then performed a manual overprivilege analysis and observed that
\overprivdChannelsAll channels exhibit overprivilege. These overprivileged channels have access to an average of \avgOverprivAPIs API functions
that they do \textit{not} need to implement the associated trigger and action functionality. We also observed that \notOverprivdChannelsAll
channels do not exhibit any overprivilege---they use all accessible APIs to implement triggers and actions.

\begin{figure}[!tb]
	\center
	\includegraphics[width=0.9\columnwidth]{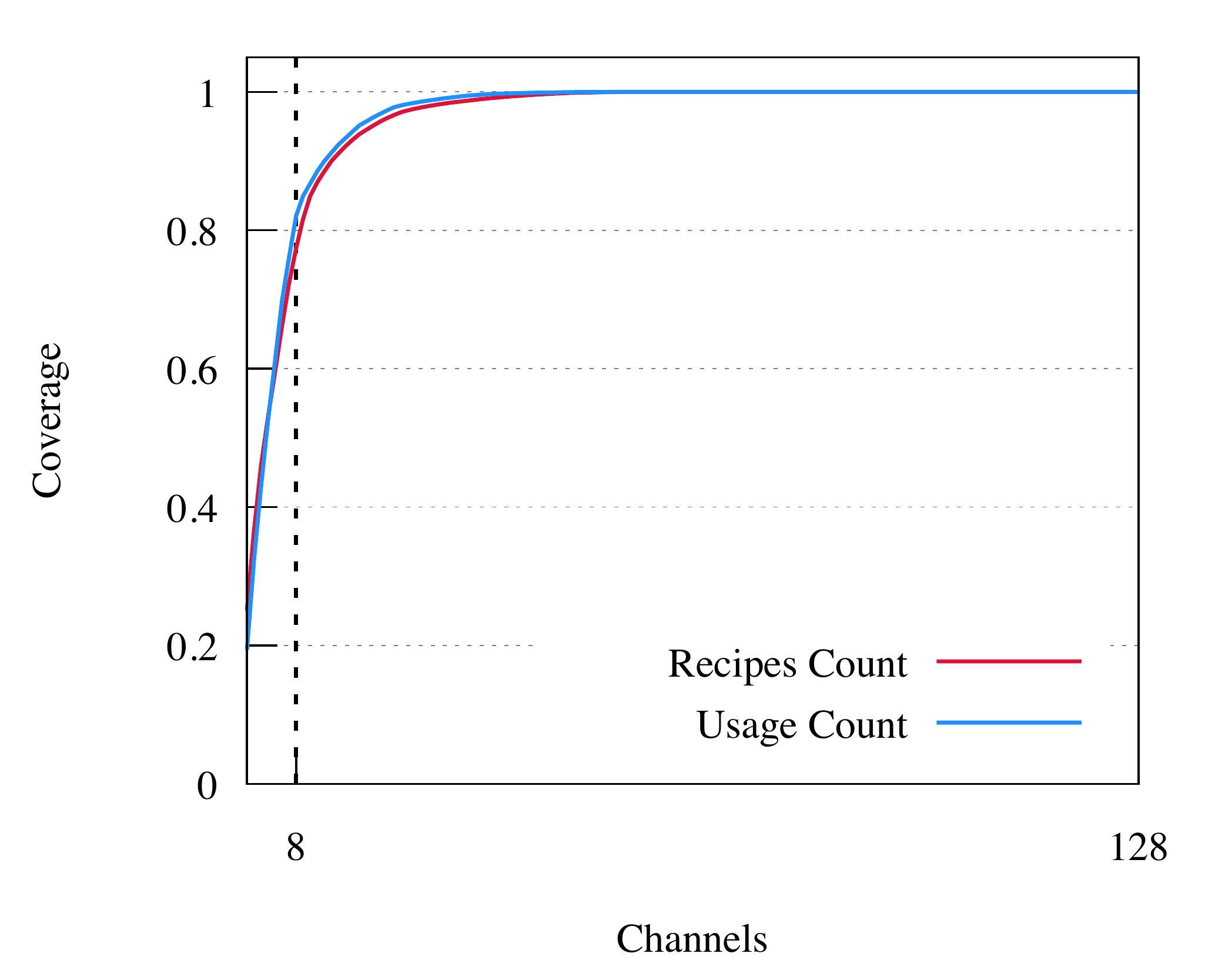}
	\caption{A CDF of our overprivilege analysis coverage. We study 8 of the top measurable channels counted in terms of the number of associated recipes, and user shares of those recipes. We also studied all $\numChannelsStudiedIoT$ cyber-physical channels that can be measured. Our overprivilege analysis covers $\recipeCoverageTrigger$ of all
  recipes that are involved in the set of channels that can be measured.}
	\label{fig:coverage}
\end{figure}

The channels we study in-depth cover $\recipeCoverageTrigger$
$(\coveredRecipesTrigger/\totalRecipesTrigger)$ of all recipes involved in the set of $\measurableChannelsAll$ measurable channels. 
Our overprivilege results potentially impact $\shareCoverageTrigger (\coveredRecipeShares/\totalRecipeShares)$
of all users of the recipes associated with the channels that can be measured.\footnote{The recipe dataset contains the number of times a recipe was shared by users~\cite{urCHITrigger}.} Figure~\ref{fig:coverage} shows 
the coverage we achieve in terms of the recipes associated with the channels we measure and in terms of the number
of users of those recipes. We use number of recipes associated with a channel and number of users of those recipes
as a metric that estimates channel adoption and popularity. We note that the coverage CDF only shows coverage
for the top 8 channels sorted in terms of recipe counts and user shares. Our actual coverage is slightly higher
than what the CDF shows since we also analyze cyber-physical channels.
Out of the total of $\numChannelsStudiedAll$ 
channels, we studied all $\numChannelsStudiedIoT$ cyber-physical channels that can be analyzed in-depth---cyber-physical channels
have the potential to create a significant security risk since they are associated with physical devices.

Based on our in-depth analysis of channel-online-service overprivilege, we revisit our examples of potentially overprivileged channels here (initially discussed in ~\xref{subsec:overprivexamples}) and confirm that the Particle and Google Drive channels have access to online service APIs that they do not need to implement their sets of triggers and actions. Appendix~A contains a complete breakdown of overprivilege by channel.

%% file: lessons.tex
\section{Lessons \& Implications of the Empirical Analysis}
\label{sec:lessons}

We highlight the lessons we extracted based on the results of our empirical analysis of IFTTT's authorization model.

\begin{itemize}

\item The channel abstraction strikes a good balance in the usability-security trade-off, but results in highly-privileged tokens residing inside IFTTT's infrastructure. Users sign in to online services only once per channel, and IFTTT obtains an OAuth token that is sufficiently privileged to execute all kinds of recipes it supports. If tokens were recipe-specific, then the user would have to sign in to the online services per recipe, thereby drastically increasing the number of prompts, leading to prompt fatigue.

\item Highly-privileged tokens inside IFTTT's infrastructure are a single point of failure, making it an attractive target for attackers. If IFTTT is compromised and becomes malicious, it can arbitrarily execute a wide variety of functions on the user's online services, as our empirical analysis shows.

\item Coarse-grained bearer tokens themselves are an attractive target for attackers and are susceptible to known OAuth attacks~\cite{oauthhack,stoakland}. We analyzed how many channels out of $\totalChannels$ are vulnerable to two types of OAuth attacks directed at the user. We found that $\numDWORAttack$ channels are vulnerable to the open redirector OAuth attack where a victim user can be tricked into logging into the authentic online service page but with an attacker-controlled redirect URI that would let the attacker obtain an OAuth bearer token. This attack is similar to that of Fernandes \etal~\cite{stoakland}. We also found that $\numDAttack$ channels are vulnerable to downgrade-only attacks. Although this attack requires a stronger assumption of a man-in-the-browser, or HTTP-only communications (no HTTPS), it still does highlight the risk that users might face due to overprivileged channels on IFTTT.

\end{itemize}

Based on the above lessons, our goal is to provide trigger-action functionality to end-users without increasing the number of OAuth permission prompts, while preventing arbitrary misuse of OAuth tokens if IFTTT is compromised. We discuss our design that achieves this goal in the next section.

\input{threatmodel}

%% file: threatmodel.tex
\noindent{\textbf{Threat Model.}} We assume that the trigger-action platform can be compromised. It can leak the OAuth bearer tokens and it can attempt to invoke actions arbitrarily. It can also try to manipulate the trigger data passing through its infrastructure. For example, if we have a recipe that saves an Instagram image to Dropbox, the untrusted trigger-action platform might instead save malware into the user's Dropbox account. A compromised trigger-action platform can leak trigger data that might be privacy sensitive. We do not prevent such leaks. Currently, the user has to trust the platform with access to data it needs to run recipes. We discuss potential ways to overcome this problem in~\xref{sec:discussion}. We assume that online services use HTTPS for their OAuth APIs,\footnote{Out of $\totalChannels$ channels, only 2 used HTTP; all others used HTTPS.} and that they are not compromised (if an online service is compromised, then an attack is possible independently of IFTTT). We consider denial of service attacks to be outside the scope.


%% file: defenses.tex
\section{\longsys Design}
\label{sec:defense}
\longsys splits the logically monolithic IFTTT architecture into a cloud service (\dcloud) that users do not trust, and several clients (\dclients). The \dcloud provides computational infrastructure to execute recipes at large scale, like IFTTT's cloud. We assume that it can be compromised by attackers. We introduce extensions to the OAuth protocol to ensure that the cloud service only has the necessary amount of privilege to execute the set of recipes of a given user. Each user must install a \dclient on a device such as a smartphone. Users connect channels to their accounts and setup trigger-action recipes with the help of the these clients. A user trusts a client to manage highly-privileged access to their online services.

We designed the OAuth protocol extensions for \sys to be open allowing anyone to implement the client portion of the protocol. Furthermore, the clients are not implemented by the same entity implementing the untrusted cloud service. Instead, we envision a community of developers building client applications and hosting them at various market places, \eg Android or Apple store. These app market models naturally result in a few well-built apps emerging, thus making it easy for users to install relatively good and secure implementations of the \dclient. Furthermore, the open source community can independently vet open source clients.

\begin{figure*}[!tb]
	\center
	\includegraphics[width=.8\textwidth]{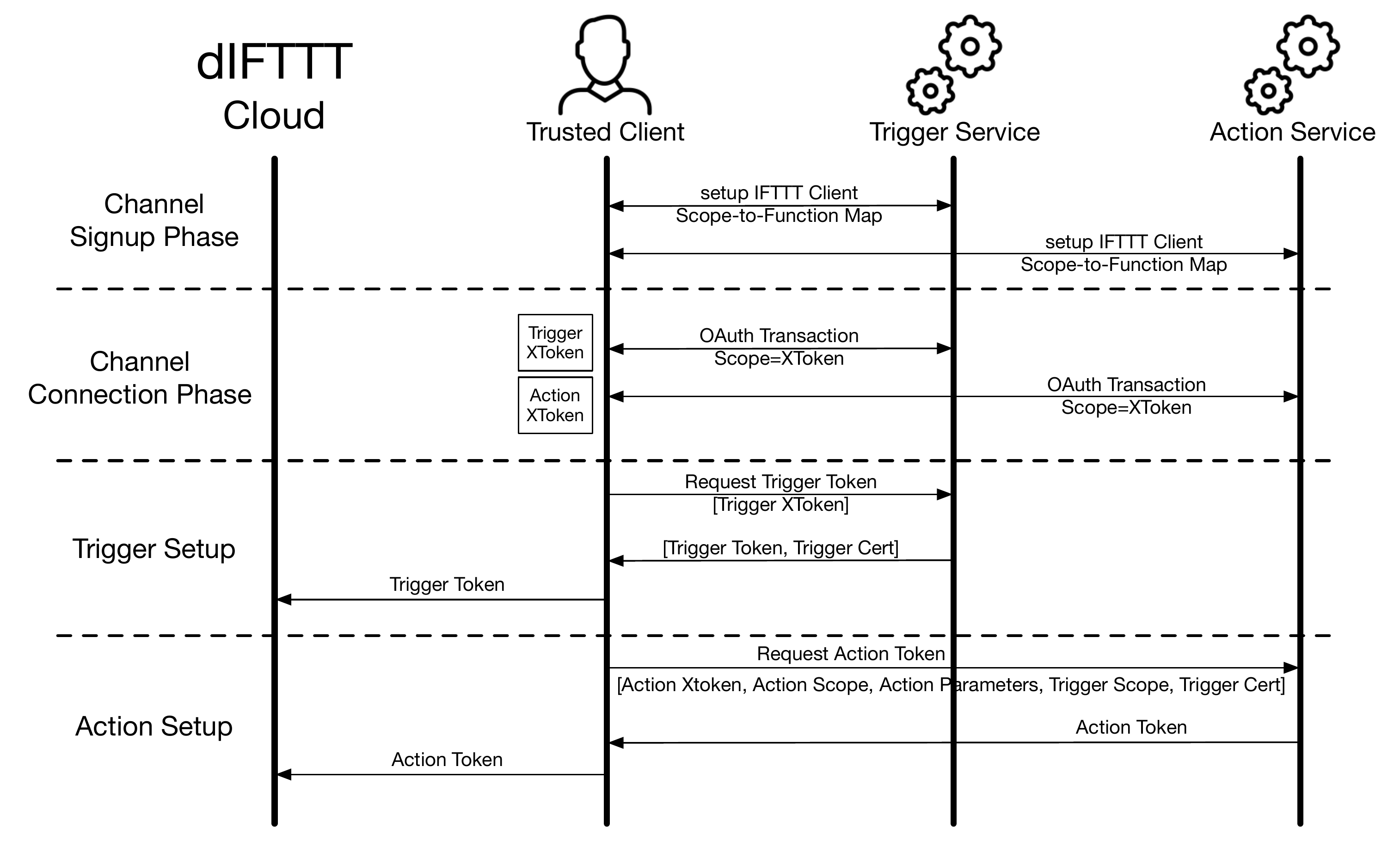}
	\caption{\sys authorization model has four phases: Channel signup phase, where the clients obtain scope-to-function maps for every online service; channel connection phase, where the clients gain XTokens to online services the user wishes to use; and trigger and action setup phases where these tokens are used to request recipe-specific tokens.}
	\label{fig:difttt}
\end{figure*}

As is the case with IFTTT, there are two phases a user must follow to create a recipe: Channel Connection, and Recipe Setup.
We will discuss how these two phases work, with the help of an example recipe taken from IFTTT:

\begin{verbatim}
IF new_item added to ShoppingList THEN 
   email new_item to x@y.com
\end{verbatim}

\noindent{\textbf{Channel Connection.}} A user will connect channels using a client. To create the above recipe, the user will have to first connect the ShoppingList and Email channels (assuming they haven't been connected before). This involves the usual step of the user logging in to the services corresponding to the channels with a username and password, and eventually accepting the OAuth scopes being requested. During this standard OAuth negotiation (we only use authorization code grant flow), the \dclient requests an \textit{XToken} (Transfer Token, see Figure~\ref{fig:difttt}). An XToken is coarse-grained and can only be used to obtain recipe-specific tokens without creating a permission prompt. As these tokens are highly-privileged (because they allow the bearer to obtain a recipe-specific token for any of the functions a particular online service provides), we encrypt their storage when they are not in main memory. The \dclient can also use a trusted-hardware-backed keystore when available on a client (\xref{sec:eval}). We introduce the notion of an XToken to maintain the usability experience of one-time authorizations of channels, and to gain the security of recipe-specific tokens.

\noindent{\textbf{Recipe Setup.}} Once the user has connected the trigger and action channels, the next step is to setup the trigger part of the recipe. This involves navigating a UI and eventually clicking on one of the trigger functions that the channel offers. In this case, \texttt{OnNewItem} is a function that fires whenever a new item is added to the user's shopping list. \dclient will treat the physical act of the user clicking a specific trigger function in the UI as an implicit authorization for it to obtain a recipe-specific token that can only execute \texttt{OnNewItem}. It transmits the XToken it obtained earlier to the trigger online service including information about the specific function for which it wants a recipe-specific token. As a return value, the trigger service will also transmit its X509 certificate to the client, in addition to the recipe-specific token (Figure~\ref{fig:difttt}).

A recipe-specific token only allows the bearer to execute a single function with specific parameters on an online service. For example, assume that the ShoppingList service offers two functions that external parties may call: \texttt{test()}, and \texttt{OnNewItem(String URL)}. The XToken allows the bearer to obtain a recipe-specific token for any of these supported functions. In our example recipe, an external party only needs to call \texttt{OnNewItem} with a String value of ``https://difttt-cloud.com/new\_item.'' Therefore, the client can obtain a recipe-specific token scoped to only execute OnNewItem(`https://difttt-cloud.com/new\_item'). That is, a scope in \sys is equivalent to the name of a function in an online service.

Our design relies on two principles to overcome the challenge of an increased number of prompts while using such fine-grained tokens: 

\begin{itemize}
\item The user authorizes the client to obtain an XToken when a channel is connected. This does not change the number of permission prompts for a user---it is the same as IFTTT. The XToken has the property of allowing the client to obtain a recipe-specific token without creating a permission prompt, as the user has already given the client that amount of privilege by authorizing it to obtain an XToken.

\item The client only uses the XToken upon an explicit user interaction. This notion is directly inspired by User-Driven Access Control~\cite{udac}.
\end{itemize}

Setting up the action part of the recipe is similar to setting up the trigger part. The user will navigate a UI and implicitly authorize the client to obtain a recipe-specific token to invoke a particular action function. However, the token exchange process is slightly different. As Figure~\ref{fig:difttt} shows, the \dclient will transmit the action XToken, the trigger service's X509 certificate, the name of the trigger function (\texttt{OnNewItem}), the action function name (\texttt{send\_email}), and any action function parameters to the action service. The action service will return a recipe-specific token and associate all of this information with that token internally, effectively tying the issued token to a particular triggering function and a particular action function.

At this point, the \dclient has obtained two recipe-specific tokens needed to execute the recipe. It transmits these tokens along with a description of the recipe to \dcloud that uses the trigger token to set up a callback to itself whenever the trigger condition (\ie new item added to shopping list) occurs.

\noindent{\textbf{Channel Signup.}} Currently, IFTTT knows which scopes to request for various trigger and action functions because channels store that scope-to-function mapping in IFTTT's infrastructure. However, in our case, this infrastructure is untrusted. \dcloud could manipulate scope-to-function mappings to trick the clients into requesting the wrong scopes. Our design solves this problem by requiring the online services to create a signed scope-to-function mapping and host those mappings at a well-known location. An online service signs its mapping using the private key corresponding to its X509 certificate. The clients retrieve these signed mappings during the channel signup phase (Figure~\ref{fig:difttt}).

\begin{figure}[!tb]
	\center
	\includegraphics[width=\columnwidth]{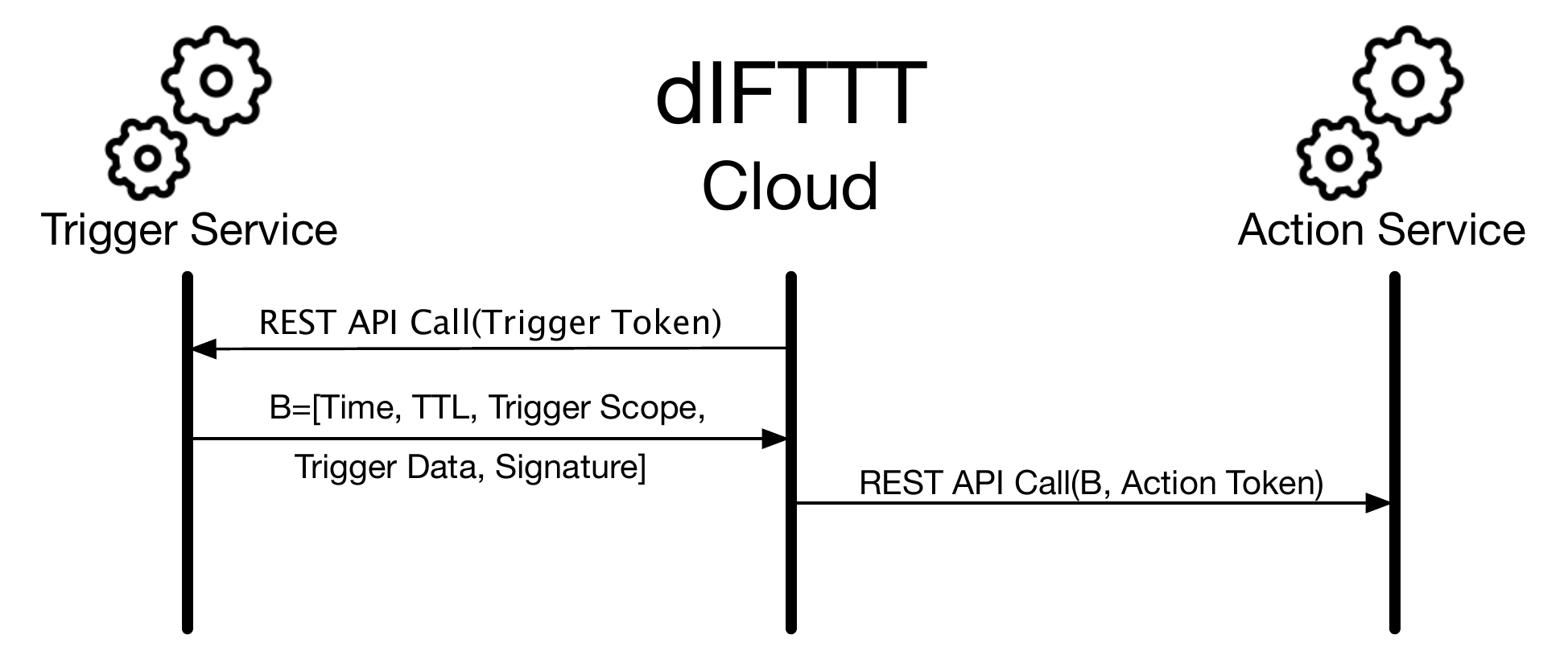}
	\caption{Recipe execution in \sys: Upon a trigger activation, the trigger service contacts \dcloud with a trigger blob. \dcloud transmits this blob and the recipe-specific action token to the action service. The trigger blob contains information the action service needs to verify that the corresponding trigger occurred.}
	\label{fig:difttt2}
\end{figure}

\noindent{\textbf{Recipe Execution.}} At runtime, whenever a new item is added to the shopping list, the trigger service will generate an HTTP call to the IFTTT cloud and pass the trigger data (in our example recipe, this will be the item that was added to the shopping list). \sys changes this process slightly, and instead requires the trigger service to generate a trigger blob (see Figure~\ref{fig:difttt2}):

$[Time, TTL, TriggerScope, TriggerData, SIG]$

where $SIG$ is a digital signature of a concatenation of the other data items. The public key of this signing private key was transmitted to the action service as part of the setup process. $Time$ is the timestamp when the blob was created, and $TTL$ specifies the period for which the blob is valid. Once the trigger service creates this blob, it will transmit it to the \dcloud. At that point, the \dcloud will lookup the appropriate recipe, and then invoke the action function using the recipe-specific token it obtained earlier.

Upon receiving the HTTP call from the \dcloud, the action service will first execute a lightweight verification process before invoking the target function. The verification steps are:

\begin{itemize}
    \item Verify that the passed recipe-specific token exists.
    \item Verify the signature on the trigger blob using the X509 certificate of the triggering service.
    \begin{itemize}
        \item Ensure that the time stamp value has increased.
        \item Verify that the Time-To-Live (TTL) value inside the trigger blob is current.
        \item Check that the trigger scope (function name) inside the blob matches what the action service was given during the setup phase.
    \end{itemize}
    \item Verify that the HTTP function being called at runtime is the same as the function name given by the trusted client to the action service during the setup phase.
    \item Finally, verify that the function parameters match those that the trusted client gave the action service during the setup phase.
\end{itemize}

If all verification checks succeed, then the action service proceeds normally and executes the \texttt{send\_email} function. We note that the recipe execution process does not depend on the \dclient, as recipe-specific tokens are already uploaded to \dcloud.

\subsection{Security Properties of our Decoupled Design} 
\label{subsec:secprop}
The above design ensures that the \dcloud can only execute user recipes whenever a trigger occurs, even if it is compromised. Here, we explain in more detail how the various components of our OAuth protocol additions and decoupled design provide this guarantee.

\noindent{\textbf{Action Misuse Resistance.}} An untrusted or compromised IFTTT cloud can invoke action functions at will, even in the absence of any triggers. Furthermore, based on our empirical study results, it could invoke a wide variety of functions given the rampant overprivilege. \sys prevents all of these problems. First, although XTokens are coarse-grained, they are never transmitted to the untrusted cloud service. Only recipe-specific tokens that can execute a single function with specific parameters are transmitted to the \dcloud. Furthermore, the \dcloud can successfully execute an action function only if it can prove that a trigger occurred within some reasonable amount of time in the past. The signed trigger blob provides this property.

\noindent{\textbf{Trigger Misuse Resistance.}} \dcloud could try to misuse the trigger blob and attempt replay attacks. However, the time stamp and time-to-live value ensures trigger blob freshness. It could also try to use a trigger blob from another trigger service or the trigger blob of a different trigger function on the same service. However, while setting up the recipe, the trusted \dclient instructs the action service to associate the name of the trigger scope (function name) with the action token. Furthermore, the signed trigger blob contains this trigger scope. Therefore, \dcloud can only use a given trigger blob for a specific action function. In other words, the \dcloud can only execute the user's recipe.

\noindent{\textbf{Trigger Data Integrity.}} The untrusted \dcloud may attempt to modify the data it receives from the triggering service before delivering it to the action service. An example of this would be a recipe that saves new images from an Instagram channel to a Dropbox account. An attacker may replace the image with malware before uploading the file to Dropbox. \sys protects against such an attack by requiring the trigger service to sign the fields of the trigger blob with its private key. When receiving the trigger blob, the action service verifies the signature using the public key that was associated with the action token during recipe setup.

\noindent{\textbf{Recipe Deletion.}} A user can delete recipes with the help of \dclient, that will issue a recipe deletion HTTP API call to the online services involved in a specific recipe. The online services will then invalidate the recipe-specific tokens. A malicious \dcloud can retain the recipe description, but it won't be able to execute any trigger or action functions because the online services will automatically refuse the HTTP calls as the tokens no longer exist.

\noindent{\textbf{No Single Point of Failure.}} Although the XToken is coarse-grained, it is never transmitted to the untrusted cloud service.  Thus, the attacker has to target and compromise individual devices to obtain the XToken. Therefore, these tokens are not a single point of failure any more.


\subsection{Usability Properties of our Decoupled Design}
\label{subsec:usableprop}

From an end-user perspective, \sys retains the concept of the one-time operation of users connecting channels to their accounts. However, as users have to use a client app, it does limit their mobility (see \xref{sec:discussion} for options to increase mobility). \sys does not add any additional OAuth prompts---it leverages User-Driven Access Control to automatically obtain the recipe-specific tokens.

As we discussed in~\xref{subsec:initobserve}, online services in general do not provide with users fine-grained control over OAuth permissions and do not provide good descriptions of the permissions being requested. However, \sys enables fine-grained control and good descriptiveness. When a \dclient requests the user's permission to obtain an XToken, it can directly list the set of online service functions for which the XToken can be used to gain access. Furthermore, the online service can provide an option for users to select the set of functions they wish to include in the XToken---\dclient will not be able to obtain recipe-specific tokens for any functions not in that set.

From a developer perspective, \sys requires changes. Specifically, it requires adding code to implement XTokens, the recipe-specific tokens, trigger blob generation, and the verification procedure. This can be a barrier to immediate adoption. However, as we discuss in~\xref{sec:eval},~\xref{sec:discussion}, we have implemented \sys in a way to ease the transition for online service developers by only requiring them to add a single annotation above HTTP methods in the server.

\subsection{Expressivity of \longsys}
\label{subsec:practicality}
For services that do not natively support a callback interface for a specific triggering condition, the trigger-action platform must poll the service and check the triggering condition itself. For example, a weather channel might only offer an API that returns the current temperature. To support a trigger that fires if the temperature goes above 80 degrees, IFTTT would poll the weather service and compute the predicate $currTemp > 80$. However, the \dcloud might simply ignore the result of the comparison, and invoke the action service repeatedly. The verification on the action end will succeed since \dcloud will obtain a valid signed trigger blob when it polls the trigger service. 

\sys handles such situations by allowing the client to associate a predicate with the action token. This predicate is expressed over fields of the trigger data part of the signed trigger blob. The \dclient simply maps the condition the user sets up while creating the recipe to a predicate and then instructs the action service to associate the predicate with the resulting recipe-specific token. At runtime, the action service performs the additional step of verifying that the predicate is true.

Encoding such stateless predicates handles a significant fraction of the kinds of conditions that IFTTT supports. We studied the triggers, actions, and online service APIs for $\numChannelsStudiedAll$ channels that covered $\recipeCoverageTrigger$ of recipes in our dataset and did not find any predicates that required storing state. We also studied the Zapier channel creation process but did not find any resources for channels to keep state~\cite{zapier}. Moreover, all Zapier predicates only involve simple boolean operators. Our prototype fully supports expressing such recipes.

%% file: eval.tex
\section{Implementation \& Evaluation}
\label{sec:eval}
We implemented \dclient on the Android platform. For additional client-side security, the \dclient will use a hardware-backed keystore, when available, to generate a key that we use to encrypt XTokens before storing them on the filesystem. Such keystores have been present in iOS devices since 2013~\cite{ios_security} and have been supported in Android devices since version 6.0~\cite{keystore}.



We built a Python library that online service developers can use to add \sys functionality. The library provides a simple annotation (\ie Python decorator) that developers can place above sensitive HTTP API methods that require recipe-specific scoping. The annotation automatically invokes the verification procedure (see ~\xref{sec:defense}). Using the Python library, we implemented the \dcloud, and two online services modeled after existing IFTTT channels: (1) an Amazon Alexa inspired ToDo list, (2) an email service. 


\subsection{Microbenchmarks}
\label{subsec:microbenchmarks}
We first quantified micro-performance factors of \sys. We created the following recipe: \textit{``IF~new\_item ==~`buy soap'~is~added~to~MyToDo~List THEN~send\_email(new\_item).''} That is, if a new ToDo item with contents ``buy soap'' is added to the list, then send an email. This recipe is representative of the kinds of recipes that users can create on IFTTT. It contains all the elements of typical recipes: a condition on data coming from the trigger service, and transfer of trigger service data to an action service function. We deployed \sys locally, created the example recipe, and then measured storage overhead, transmission overhead, and developer effort.\footnote{For microbenchmarks, deployment location does not affect the quantities under study.} We found that using \sys imposes negligible overhead: Each recipe requires an additional $3.5KB$ in terms of storage, and an additional $7.5KB$ of transmission per execution. Online service developers using our prototype library only need to add a single line of code per HTTP API function---this is the same as that required by the popular oauthlib library for Python. We elaborate on the results below. 

\noindent\textbf{Storage Overhead.} Using \sys requires online services to store additional state: An online service needs to store an XToken for each trusted client that allows the client  to create fine-grained tokens for individual recipes. The online service also needs to store \sys fine-grained tokens for each recipe. These tokens include additional fields (\eg time, TTL), so they impose storage overhead on the online service. We computed the required storage for the baseline IFTTT system, and for \sys. Our results show that each \sys recipe creates a $3.5KB$ overhead in addition to the $0.8KB$ required to store the XToken, compared to the $0.8KB$ storage cost for the baseline IFTTT system. This extra token storage cost is negligible given the low price of storage and quantity of other user data that these systems collect. 

\begin{figure}[!tb]
	\center
	\includegraphics[width=\columnwidth]{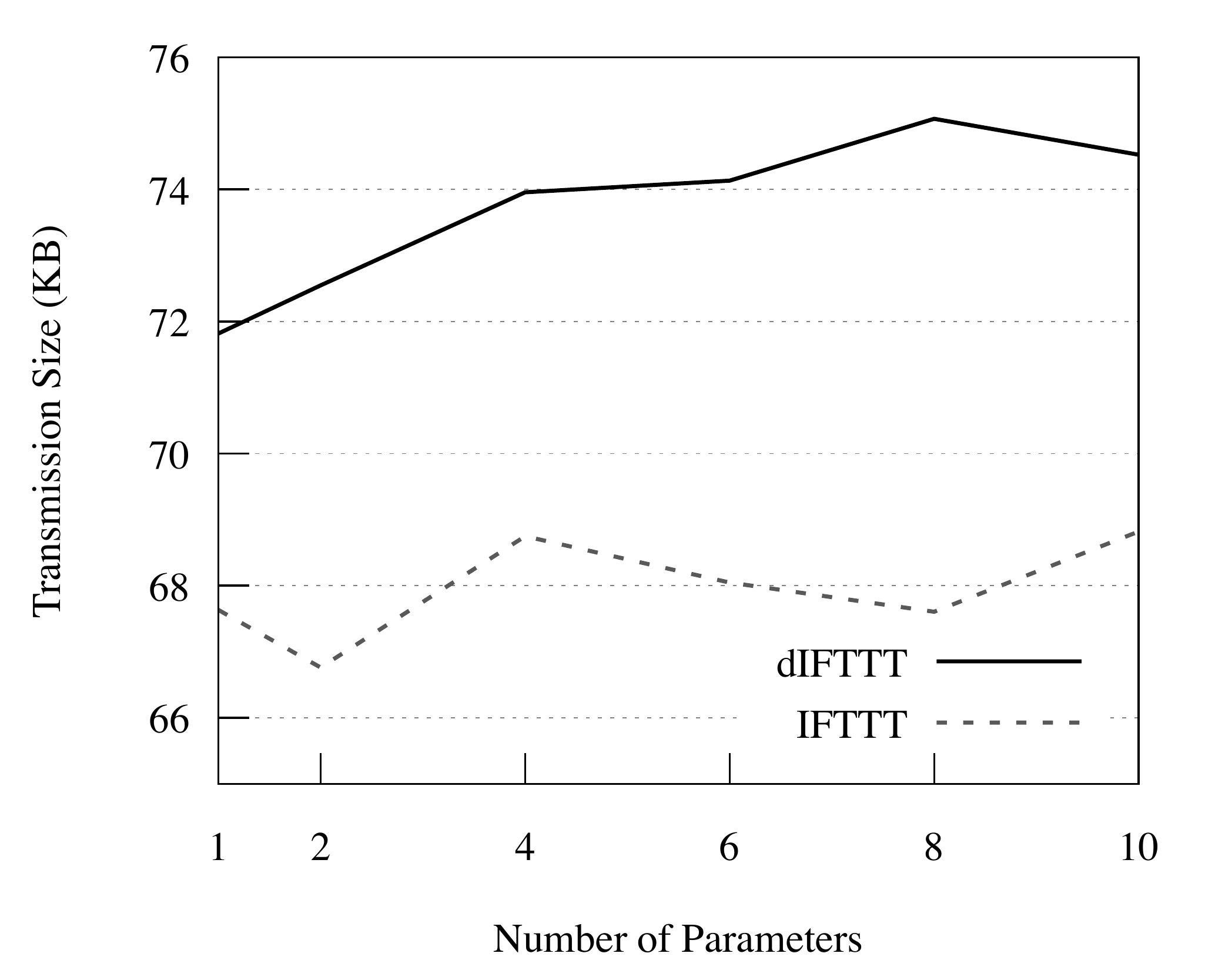}
	\caption{Average total transmission size of IFTTT and dIFTTT for $1-10$ parameters for $5$ experiments. Although there is a linear increasing trend in both systems, the difference among the two remains negligible.\vspace{-2em}}
	\label{fig:transmission}
\end{figure}

\noindent\textbf{Transmission Overhead.} Executing a recipe on \sys requires transmitting more data over the network. This overhead is the result of additional data in the trigger blob (Figure~\ref{fig:difttt2}) including time, TTL, and sign. To evaluate the transmission overhead, we computed the transmission size of recipe execution in the baseline case and compared it to the same quantity in the \sys case. We varied the number of function parameters passed ($1-10$) and present the average result of five experiments. The number of function parameters matters because the recipe-specific token information encodes data about the specific function being executed. We used Wireshark~\cite{wireshark} to measure the flow sizes associated with ports assigned to online services and the \dcloud. Figure~\ref{fig:transmission} presents this overhead for different number of function parameters for the two systems. In our experiments, \sys created $6-11\%$ overhead. Even when using $10$ parameters the transmission overhead does not exceed $7.5KB$.  


\noindent\textbf{Developer Effort.}
We developed \sys as a library for trigger and action services to make it easy for online service developers to transition to the \sys model. Developers must only add a single additional line of code per function to protect it with \sys verifications. When compared to existing OAuth libraries, such as the popular oauthlib~\cite{oauthlib}, this is the same amount of effort---developers using oauthlib must also place a single annotation above HTTP API methods to create scopes.

\subsection{Macrobenchmarks}
\label{subsec:macrobenchmarks}
We measured end-to-end latency and throughput of recipe execution. We hosted the \dcloud and two online services on separate Amazon t2.micro EC2 instances. Each instance was configured with one 64-bit Intel Xeon Family vCPU@2.5 GHz, 1GB memory, 8GB SSD storage, Ubuntu 14.04 with Apache2, and MySQL Server 5.5. Our results show a modest $15ms$ latency increase, and $2.5\%$ throughput drop in the online service when compared to the baseline (online service with no \sys protections). This does not represent an inhibiting overhead for an online service especially when considering the effect of network latency and the lack of real-time requirements in these systems. We used the same ToDo list recipe for our tests.

\begin{figure}[!tb]
	\center
	\includegraphics[width=\columnwidth]{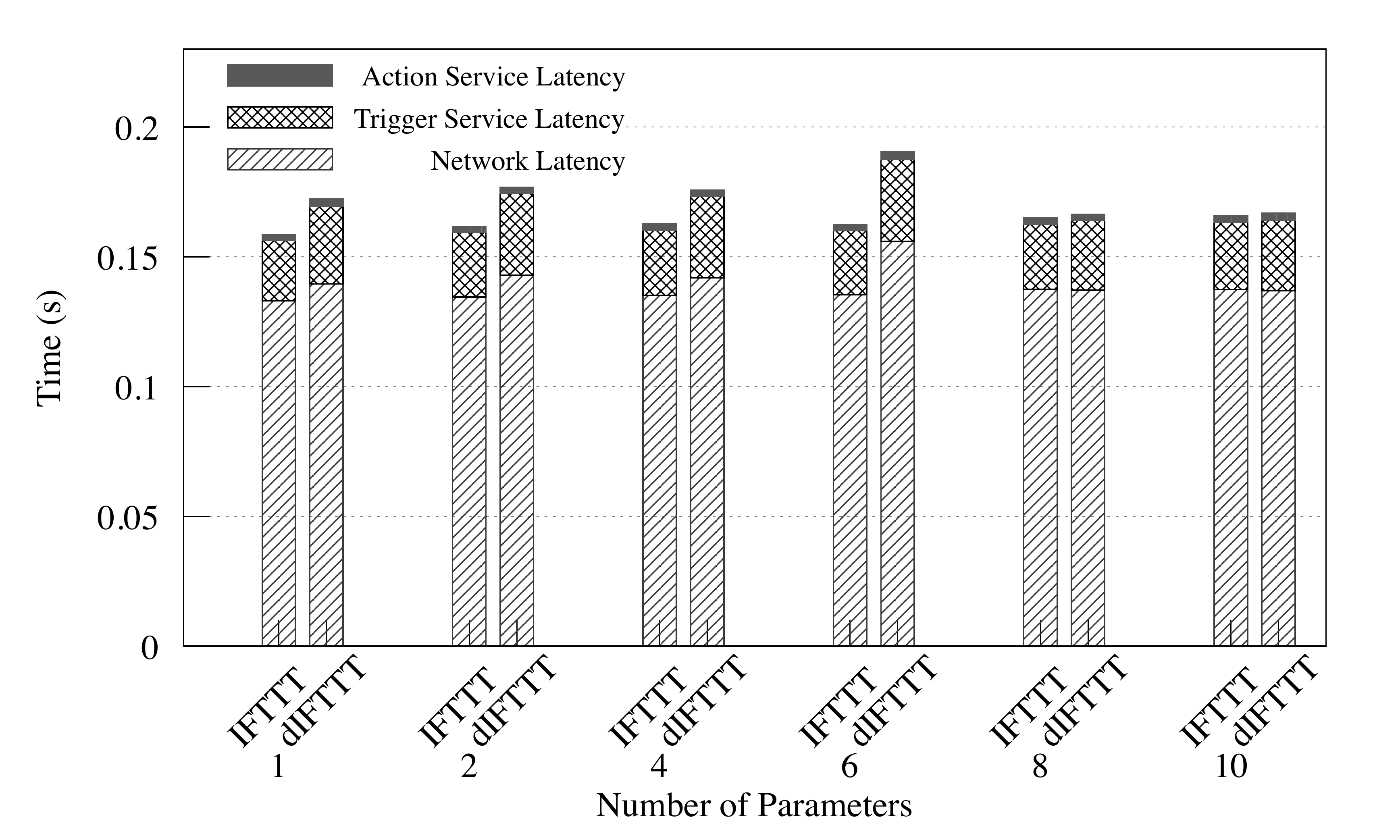}
	\caption{\sys adds less than $15ms$ of verification latency to recipe execution when compared to the baseline IFTTT case.}
	\label{fig:latency}
\end{figure}

\noindent\textbf{End-to-End Latency.} We measured the time between the trigger service being activated due to an item being added to our ToDo list example recipe, and the time the action service issues a \texttt{send\_email} call. This time includes network latency, the time to generate a signed trigger blob, and the time to verify the trigger blob and the action token, in the case of \sys. Our baseline case is the IFTTT system, and it only includes network latency, and time to execute the trigger and action functions without any \sys verification. We varied the number of function parameters on the action service between $1$ and $10$. Figure~\ref{fig:latency} presents the results of these experiments. Our results show that excluding the network latency, the maximum verification overhead is less than $15ms$. For typical recipes, that send emails, SMSs, or invoke actions on physical devices over a network, we consider this additional latency to be acceptable.

\noindent{\textbf{Throughput.}} We measure throughput as the number of recipes executed per second, under a load of 2000 concurrent HTTP requests. We computed this concurrency level by examining the number of times the most popular IFTTT channel was used in recipes (IF Notification channel was used in $1,514,188$ recipes in our dataset). As per IFTTT's documentation, this channel will contact an online service once every 15 minutes~\cite{ifttt15min}, meaning that an online service would receive approximately $1,682$ $requests/second$. Therefore, we chose $2000$ as an upper-bound for the number of concurrent requests a service would have to process. We used ApacheBench~\cite{ab} to conduct throughput testing of \sys and IFTTT by sending $10,000$ trigger activations with upto $2000$ concurrent activations at a time. Table~\ref{tab:throughput} presents our results, averaged over three separate runs. We find that \sys decreases throughput by only $2.5\%$.

\begin{table}[tb!]
	\center
	\begin{tabular}{c c c c c }
		\toprule
		 	 & \multicolumn{2}{c}{\textbf{\sys}} & \multicolumn{2}{c}{\textbf{IFTTT}} \\
		 	 & \texttt{Avg} & \texttt{SD} & \texttt{Avg} & \texttt{SD} \\
		\toprule
		\textbf{Throughput (req/sec)} & 94.03 & 8.48 & 96.46 & 5.74\\ \bottomrule
	\end{tabular}
	\caption{\sys reduces throughput by $2.5\%$ when compared to IFTTT. We used ApacheBench to send $10,000$ trigger activations with upto $2000$ concurrent activations at a time.\vspace{-2em}}
	\label{tab:throughput}
\end{table}

%% file: discussion.tex
\section{Discussion}
\label{sec:discussion}


\noindent{\textbf{Transitioning to \sys.}} Our design has the limitation of requiring changes in online services to support the recipe-specific tokens and the cryptographic extensions to the OAuth 2.0 protocol. This can be a barrier to immediate adoption. We ease this transition by modeling our implementation after the popular oauthlib~\cite{oauthlib} where developers only have to add a single-line annotation above HTTP methods in the server that need to be protected by recipe-specific tokens. A short-term option is to also construct a trusted proxy or shim around online services that adds \sys support. In this case, users would login to the proxy instead of the online service, and only the proxy is \sys-aware. Once online services have added \sys support, the trusted proxy can be removed.

\noindent{\textbf{\dclient use.}} In IFTTT, users can login to the IFTTT website and create recipes from any client device. However, \sys requires users to create recipes via a client device they trust (e.g., their smartphone), which stores XTokens in a private file system. Although our current client prototype does not support transferring client state from one device to another, building such functionality is fairly straightforward. One possible solution is to provide an export function to save the current client state to a disk image, and then provide an import function to load that client state into another device. If a client device is lost, then user recipes continue to execute normally. However, the user will have to download the client again on another device, and go through the channel connection phase to re-establish the XTokens to create future recipes.

\noindent{\textbf{Other potential solutions to reduce the negative impact of overprivilege.}} One option is for online services to issue OAuth tokens that must be refreshed frequently. If the trigger-action platform is compromised, the online services can simply stop processing refresh requests from IFTTT. This technique reduces the useful attack window to the refresh interval plus the time it takes for the knowledge that the platform was compromised to propagate to the online services. This solution requires detecting a compromise in a timely fashion. Furthermore, such a design does not reduce the privilege of the trigger-action platform---it still remains an attractive target.


Another solution is to use OAuth 1.0 tokens because these are not immediately useful to attackers if they are stolen in isolation. It requires stealing the shared signing secret as well. However, if the trigger-action platform is compromised, then the attacker gains access to the signing key as well. 

\noindent{\textbf{Client-Device Loss.}}
If a client device is lost, existing procedures to erase device data take care of removing OAuth tokens. Also, an ``erasure-app'' can be built to automatically contact online services and invalidate tokens with co-operation from our modified OAuth helper library. We leave implementing this to future work. 

\noindent{\textbf{Data confidentiality.}} Our design currently reduces the privilege of the \dcloud---it only gains access to APIs and hence data it needs to run the user's recipes. This is an improvement over the current state-of-the-art where we have shown through our empirical analysis that an attacker can gain wide access to data and devices. However, even with our improvements, an attacker can still gain access to sensitive information simply by passively recording recipe execution. A potential way to provide data confidentiality in this case is to encrypt data passing through the \dcloud. However, this can result in a loss of expressivity. Currently, IFTTT can evaluate predicates on trigger data (see our weather data example in~\xref{subsec:practicality}). Although the action service can solely evaluate these predicates, it does increase computational burden, thus defeating the purpose of a system like IFTTT. As our analysis shows, the predicates are stateless and involve simple comparison operators. Therefore, a potential solution is to leverage advancements in use-case-specific homomorphic encryption for secure integer comparison, rule matching, etc., to allow the least-privilege \dcloud to evaluate predicates on encrypted data~\cite{seal,blindbox}. We leave this to future work. 

%% file: related.tex
\section{Related Work}
\label{sec:related}

\noindent\textbf{Trigger-Action Platform Studies.} A few studies have investigated IFTTT in recent years, although in different contexts. Ur \etal~\cite{urCHITrigger} crawled the site in 2015, 
collecting 224,590 IFTTT programs shared by over 100,000 different users. 
Their study shows many interesting statistics including the number of 
different trigger and action channels used by IFTTT users. In contrast, we conduct an empirical overprivilege analysis of how channels interact with the corresponding online services.

Poirot is a security analysis tool that finds vulnerabilities in systems that occur due to discrepancies between a designer's view of a system and that of an attacker~\cite{fse16}. The authors apply Poirot to IFTTT as a case study and find a previously unknown login CSRF based attack where data from one user account can be written to another unrelated attacker account. In contrast, we assume that the IFTTT platform can be compromised, and introduce a decoupled design where the cloud component executes recipes at scale and is untrusted. \sys ensures that if the cloud component is compromised, the attacker cannot arbitrarily invoke actions. Instead, it can only invoke actions if it can prove trigger occurrence.

TrigGen is a tool that aims to avoid errors caused by users incorrectly creating rules that have insufficient triggering conditions~\cite{plas16}. Our work is not focused on recipe correctness. We study the security of the way that IFTTT interacts with online services.

\noindent\textbf{OAuth Security Analyses.} Since the Open standard for Authorization (OAuth) debuted 
in 2007~\cite{oauth1}, a number of studies discovered flaws in the protocol 
and the way the protocol was implemented in web sites~\cite{fett-oauth2,oauthAdvisory:2009-1,homakov-1,homakov-2,breakSEcurity-1,breakSEcurity-2,sun12ccs,wang12oakland,wang13ccs,wang13sec}. 
Nonetheless, the OAuth protocol is still popular and it is now commonly used 
in mobile applications as well. Since the protocol was initially 
designed for web sites, some of the important details of the protocol 
was up to developers' interpretation when adapting OAuth to a mobile 
application. 

Recent work scrutinized implementations of OAuth in many Android 
mobile applications~\cite{oauthhack,wang15acsac,shehab14icms}, 
showing that the majority of implementations 
were vulnerable~\cite{oauthhack,wang15acsac}. 
In the context of analyzing the security of the SmartThings smart home 
app platform, Fernandes \etal~\cite{stoakland} showed that a third-party 
Android application embedded a client ID and the secret needed to 
authenticate the relying party in the application binary, allowing an 
impersonation attack and injecting a backdoor pin code to 
a doorlock as a result. 

Our work is an addition to this growing list of work discovering 
vulnerabilities associated with implementing the OAuth protocol. 
However, our focus is to understand overprivilege granted 
to IFTTT independently of an online service implementing 
the protocol securely, and then design defenses. The OAuth related vulnerability 
that we discovered for several online services simply makes 
the attack easier and more widely applicable. Beside vulnerabilities in implementation, other attacks on trigger-action platforms may also expose user data to attackers. Massive data leaks are happening with increasing commonality. Target~\cite{target-leak}, Ashley Madison~\cite{ashmad-leak}, and US voters database~\cite{voter-leak} are some of the most recent examples of such high profile leaks. Our work introduces the first decoupled trigger-action platform design with the security property of only allowing the attacker who compromises the platform to execute specific user recipes.

Fett \etal conducted a formal security analysis of the OAuth 2.0 standard, and in the process discovered new vulnerabilities~\cite{formaloauth}. They also propose fixes and then prove the security of the protocol in an expressive web model. These contributions are orthogonal to ours and our work will benefit from their fixes to the OAuth protocol.

\noindent\textbf{Cloud Platform Compromise.} Beside vulnerabilities in OAuth implementation, other attacks on trigger-action platforms may also expose user data to attackers. Massive data leaks are commonplace. Target~\cite{target-leak}, Ashley Madison~\cite{ashmad-leak}, and US voters database~\cite{voter-leak} are some of the most recent examples of such high profile leaks. Our work introduces the first decentralized trigger-action platform design with the security property of only allowing the attacker who compromises the platform to execute specific user rules.

%% file: conclusion.tex
\section{Conclusions}
\label{sec:conclusions}

Trigger-Action platforms enable users to stitch together various online services that represent data and physical devices to achieve useful automation. These platforms work by gaining privilege to access user data and devices in the form of OAuth tokens.  These systems pose a long-term security risk---if they are ever compromised, attackers can use these tokens to \textit{arbitrarily} manipulate data and devices. In this work, we performed the first measurement study aimed at quantifying the risk users face in the event of a compromise. We studied the authorization model of If-This-Then-That (IFTTT), a platform with wide support for user data and devices with an active and large user community. Using semi-automated measurement tools that we built, we analyzed $\numChannelsStudiedAll$ channels, including $\numChannelsStudiedIoT$ cyber-physical channels, and achieved a coverage of $\recipeCoverageTrigger$ of all recipes associated with the set of $\measurableChannelsAll$ measurable channels. We found that $\overprivdChannelsAll/\numChannelsStudiedAll$ channels have access to online service APIs that they do \textit{not} need to implement their triggers and actions. To demonstrate the abilities of an attacker, we used overprivileged tokens to reprogram a Particle chip's firmware and delete a user's Google Drive files. More generally, we conclude that attackers can misuse tokens to arbitrarily manipulate devices and data in current trigger-action platforms.

Motivated by these findings, we designed, built and evaluated \sys, the first decoupled trigger-action platform that provides trigger-action functionality without the corresponding long-term security risks.  \sys splits the logically monolithic IFTTT architecture into an untrusted cloud service and a set of clients for users. We introduced the concept of recipe-specific tokens that, upon verification, guarantee that the recipe was executed correctly on valid trigger inputs. Recipe-specific tokens guarantee that even in the event of a total compromise of the cloud service, it cannot cause unauthorized actions to be executed on an action channel.  We also introduced the notion of the Transfer Token (XToken), and apply it to achieve the security of recipe-specific tokens without increasing the number of authorization prompts for users, when compared to IFTTT.  We built a Python library that online service developers can use to add \sys support with a single-line annotation.  We conducted a range of micro- and macro-benchmarks to establish that \sys poses modest overhead: an additional $15ms$ latency in executing a recipe end-to-end, and a $2.5\%$ throughput drop while servicing 2000 concurrent trigger activations.



%% file: ack.tex
\section*{Acknowledgements}
This material is based in part upon work supported by the National Science Foundation under Grant No. 1318722.

%% file: appa.tex
\section*{Appendix A: An In-depth look at overprivilege}
\label{sec:appa}

Table~\ref{overprivresults} shows a detailed breakdown for the in-depth measurement study of overprivilege. Table~\ref{tab:overprivdetail} shows example overprivileged APIs that IFTTT can access.

\begin{figure}[!tb]
\center
\includegraphics[width=\columnwidth]{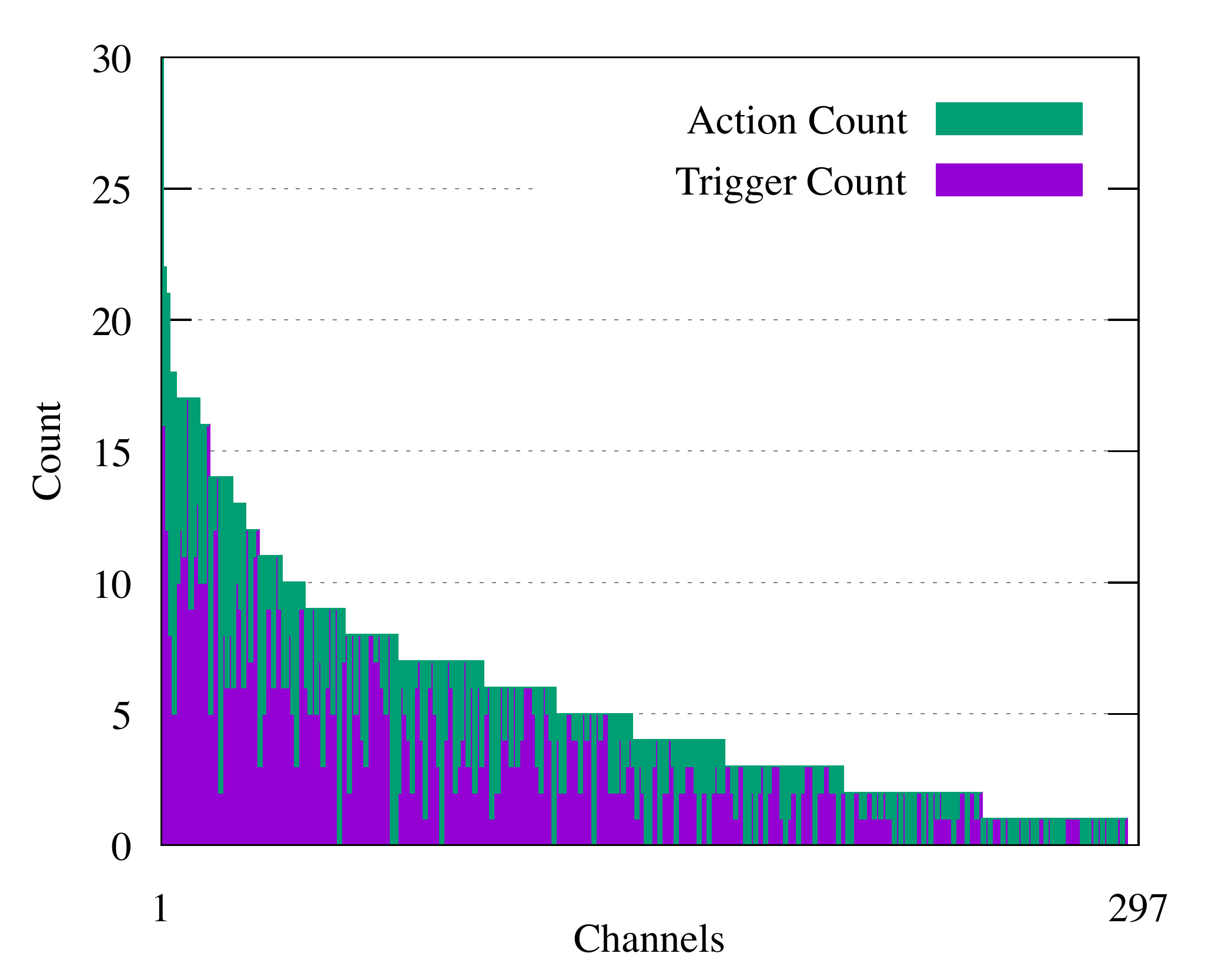}
\caption{Number of triggers and actions per channel sorted by their count. A channel has on average $5.5$ triggers and actions.}
\label{fig:trigactDist}
\end{figure}

\input{overprivimplic}

\section*{Appendix B: Distribution of triggers and actions in IFTTT channels}
\label{sec:appb}

Figure~\ref{fig:trigactDist} shows a distribution of the number of triggers and actions per channel in IFTTT.

\begin{table*}[t]
\center
\begin{tabular}{ c  | c c c c c c c c}
\toprule
~ & \textbf{Channel} & \textbf{Triggers} & \textbf{Actions} & \textbf{Total APIs} & \textbf{Overpriv. APIs} & \textbf{Not overpriv. APIs} & \textbf{Unknown Access} & \textbf{Measurement Strategy} \\
\toprule
\multirow{8}{*}{\rotatebox[origin=c]{90}{Non-Cyber-Physical Channels}} 
&Facebook & 10 &  3 & 217 & 172 & 18 &  27 &  scope-mirroring\\ \cmidrule{2-9}
&Twitter & 10 &  6 & 92 & 72 & 16  & 4 & scope-mirroring \\ \cmidrule{2-9}
&Dropbox & 2 & 3 & 81 & 61 & 16 &  4 & scope-mirroring \\ \cmidrule{2-9}
&Gmail & 6 & 1 & 56 & 37 & 7 & 12  & scope-mirroring \\ \cmidrule{2-9}
& Google Calendar & 3 & 1 & 37 & 33  & 4 & 0 & scope-mirroring \\ \cmidrule{2-9}
&YouTube & 3 & 0 & 50 & 31 & 5 & 14 &  scope-mirroring \\ \cmidrule{2-9}
&Google Drive &  0 & 4 & 34 & 31 & 3 & 0 & scope-mirroring \\ \cmidrule{2-9}
 &Pocket &  4 & 1 & 13 & 9 &  4 & 0 & scope-mirroring \\ \toprule
\multirow{16}{*}{\rotatebox[origin=c]{90}{Cyber-Physical Channels}}
&MyFoxHomeControl &  5 & 4 & 72 & 61 & 9 & 2 & scope-mirroring \\ \cmidrule{2-9}
&ATT M2X &  3 & 3 & 93 & 34 &  12  & 47 &  scope-mirroring \\  \cmidrule{2-9}
&Google Glass &  0 & 1 & 22 & 19 & 2 & 1 & downgrade \\ \cmidrule{2-9}
&Camio & 3 & 4 & 27 & 17 & 3 & 7 & scope-mirroring \\ \cmidrule{2-9}
&Particle &  4 & 2 & 34 & 15 & 6 & 13 &  actual-ifttt-token \\ \cmidrule{2-9}
&Recon & 0 & 1 & 21 & 15 & 0 & 6 & scope-mirroring \\ \cmidrule{2-9}
&UPByJawbone & 13  & 4 & 46 & 11 & 12 &  23 &  scope-mirroring \\ \cmidrule{2-9}
&Ubi & 1 & 1 & 6 & 3 & 3 & 0 & downgrade-open-redirect \\ \cmidrule{2-9}
&Pavlok &  0 & 4 & 20 & 2 &  9 &  9 &  downgrade \\ \cmidrule{2-9}
&Netatmo Welcome & 9 & 0 & 8 & 2 & 3 & 3 & downgrade-open-redirect \\ \cmidrule{2-9}
& {Netatmo Thermostat} &  6 & 8 & 6 & 0 & 5 & 1 & scope-mirroring \\ \cmidrule{2-9} 
&{Netatmo Weather Station} & 17 &  0 & 2 & 0 & 1 & 1 & scope-mirroring \\ \cmidrule{2-9}
&{Nest Thermostat} & 4 & 3 & 1 & 0 & 1 & 0 & scope-mirroring \\ \cmidrule{2-9}
&{Nest Protect}  & 5 & 0 & 1 & 0 & 1 & 0 & scope-mirroring \\ \cmidrule{2-9}
&{Nest Cam} &  3 & 0 & 1 & 0 & 1 & 0 & scope-mirroring \\ \cmidrule{2-9}
&{LaMetric} &  1 & 1 & 1 & 0 & 1 & 0 & downgrade-open-redirect \\
\bottomrule
\end{tabular}
\caption{Channel-Online-Service overprivilege results detail. Only $6$ of the $\numChannelsStudiedAll$ did not have overprivileged access to online service APIs. \texttt{actual-ifttt-token} means that we were able to obtain the token that IFTTT itself was using through an administrative API.}
\label{overprivresults}
\end{table*}

%% file: overprivimplic.tex
%

\begin{table*}[t]
\center
\begin{tabular}{c p{11cm} p{4cm}}
\toprule
\textbf{Channel} & \textbf{Example Overprivileged APIs} & \textbf{Description} \\
\toprule
Facebook & \url{https://graph.facebook.com/v2.7/from.id_status-id} & Updates status that was posted by the app itself \\ \cmidrule{2-3}
  & \url{https://graph.facebook.com/v2.7/object-id/likes} & Deletes likes \\ \cmidrule{2-3}
  & \url{https://graph.facebook.com/v2.7/payment-id/refunds} WITH BODY currency=USD\&amount=value & Initiates a refund \\ \toprule

Twitter & \url{https://api.twitter.com/1.1/statuses/destroy/240854986559455234.json} & Deletes a status message  \\ \cmidrule{2-3}
        & \url{https://api.twitter.com/1.1/statuses/retweet/241259202004267009.json} & Re-tweets a message \\ \cmidrule{2-3}
        & \url{https://api.twitter.com/1.1/account/update_profile_banner.json?width=1500&height=500&offset_top=0&offset_left=0&banner=FILE_DATA} & Updates profile banner \\ \toprule

Dropbox & \url{https://api.dropboxapi.com/2/files/delete} & Deletes a file \\ \cmidrule{2-3}
        & \url{https://api.dropboxapi.com/2/sharing/add_file_member} & Shares a file with a member \\ \cmidrule{2-3}
        & \url{https://api.dropboxapi.com/2/sharing/create_shared_link} & Creates a file sharing link \\ \toprule

Google Drive & \url{https://www.googleapis.com/drive/v3/files/file-id} & Deletes a file \\ \cmidrule{2-3}
             & \url{https://www.googleapis.com/drive/v3/files/file-id/permissions} & Creates a permission for a file \\ \cmidrule{2-3}
             & \url{https://www.googleapis.com/drive/v3/files/file-id/revisions/rev-id} & Permanently deletes a revision of a file \\ \toprule

Particle & \url{https://api.particle.io/v1/devices/device-id} & Flashes a device with a pre-compiled binary \\ \cmidrule{2-3}
         & \url{https://api.particle.io/v1/devices/device-id} & Unclaims a device \\ \cmidrule{2-3}
         & \url{https://api.particle.io/v1/devices/device-id} WITH BODY name=new\_name & Renames a device \\ \toprule

MyFox Home Control & \url{https://api.myfox.me:443/v2/site/site-id/device/cam-id/camera/recording/stop} & Stops camera recording \\ \cmidrule{2-3}
        & \url{https://api.myfox.me:443/v2/site/site-id/device/dev-id/heater/on} & Sets heater to `on' mode \\ \cmidrule{2-3}
        & \url{https://api.myfox.me:443/v2/site/site-id/device/dev-id/socket/on} or /off & Turns a device on or off \\ \toprule

Ubi & \url{https://portal.theubi.com/v2/ubi/list} & Lists all connected Ubi devices \\ \cmidrule{2-3} 
    & \url{https://portal.theubi.com/v2/ubi/ubi-id/speech?phrase=hello} & Sends a speech command \\ \cmidrule{2-3}
    & \url{https://portal.theubi.com/v2/ubi/ubi-id/location} & Returns geo-location of the Ubi \\ \toprule

Google Glass & \url{https://www.googleapis.com/mirror/v1/locations} & Gets a location that is associated with a timeline item \\ \cmidrule{2-3}
  & \url{https://www.googleapis.com/mirror/v1/timeline/id} & Deletes a timeline item \\ \cmidrule{2-3}
  & \url{https://www.googleapis.com/mirror/v1/contacts} & Gets all contacts \\

\bottomrule
\end{tabular}
\caption{Examples of overprivileged APIs channels can access that are not used in any of their triggers or actions.}
\label{tab:overprivdetail}
\end{table*}

\noindent \textbf{Particle.} Our API testing reveals that the Particle
IFTTT channel has the ability to flash new firmware to a chip. We used a token with \texttt{scope=ifttt}, which is identical to what 
the Particle IFTTT channel requests, and reprogrammed a chip by simply using a REST API call to confirm this. This can
completely change the functionality of the Particle chips and cause a variety of security and safety issues if the
corresponding token is stolen. We also observe that
the Particle OAuth prompt only provides the user with a binary choice of either approving or denying the permission request.

\noindent \textbf{Google Drive.} Our API testing reveals that the Google Drive IFTTT channel has the ability
to delete a user's files. We confirmed this behavior by using a token with the same scope as what the Google Drive IFTTT
channel requests. This can cause data loss if the corresponding token is stolen. We observe that the Google Drive
channel requests multiple scopes. However, the OAuth prompt only provides the user with a binary choice of either approving
or denying the request.

We also study the following channels' overprivilege results in detail, in addition to the above:

\noindent \textbf{Facebook.} This channel can post status messages and upload photos. However, our API testing
reveals that the channel has overprivileged access to the Facebook API that allows it to delete likes on various types
of objects and also initiate refunds. This can lead to potential financial issues. This channel requests
relatively fine-grained scopes, and the corresponding OAuth permissions prompt allows users to modify
the requested permissions.

\noindent \textbf{Twitter.} This channel can post tweets on behalf of the user, can send direct messages, and update
the biography page. However, our API testing reveals that the channel has overprivileged access to the Twitter API
that allows it to delete tweets, retweet other tweets, delete direct messages, follow friends, delete friends, and
even update the profile banner. This can cause data loss and even profile defacement if the corresponding token in stolen.
We also observe that Twitter's REST API is based on OAuth 1.0 and therefore there
is no scope argument to the authorization URL. The permissions are set up in the developer app console. Furthermore,
the Twitter OAuth prompt only provides a binary choice to the user while requesting authorization---either approve
all requested permissions, or deny the request.

\noindent \textbf{Dropbox.} This channel provides actions to add new files to a user's account or to append
to existing files. However, our API testing reveals that the channel has overprivileged access to the Dropbox
API that allows it to delete files, change file sharing settings, and even create file sharing links. This can
cause data loss and leakages if the corresponding token is stolen. We also
observe that the Dropbox channel does not use a scope argument in the OAuth authorization URL. The permissions
are set in the developer app console. Furthermore, the Dropbox OAuth prompt only provides the user with a binary choice---either
approve all requested permissions, or deny the request with no way to customize the granted permissions.

\noindent \textbf{MyFox Home Control.} This channel can arm or disarm the MyFox security system. However, our API
testing reveals that the channel has overprivileged access to the MyFox Home Control API that allows it to stop
live video recording, turn on/off electric devices, and change the state of the heaters. This can result in
security breaches, overheating and large utility bills if the corresponding token is stolen. We also observe that
MyFox Home Control does not provide any kind of scoped access. This forces the channel to request complete
access to the API. Furthermore, the MyFox Home Control OAuth prompt only provides a binary choice during authorization---either approve
all requested permissions, or deny the request.

\noindent \textbf{Ubi.} This smart home channel can trigger whenever there is spoken command and it can speak announcements.
However, our API testing reveals that the channel has overprivileged access to the Ubi API that allows it to
also send speech commands, get the geo-location of the device, and get a list of all authenticated Ubis for a user.
These overprivileged APIs can enable an attacker to send commands to the Ubi device causing security and safety issues.
We find that this online service provides the opaque \texttt{scope=ifttt} and provides an all-or-nothing choice to the
user when a channel requests privilege. Furthermore, the Ubi online service is vulnerable to the downgrade-with-open-redirect attack.

\noindent \textbf{Google Glass.} This popular wearable-category channel can send a notification to the Glass device.
However, our API testing reveals that the channel has overprivileged access to the Glass API that allows it to
get all user contacts, delete timeline items, and get locations associated with timeline items. This can result
in data loss or data leakage. We find that this online service provides fine-grained scopes, but surfaces an all-or-nothing
OAuth prompt to the user. 

